\documentclass[aps, prb, reprint, groupedaddress, showpacs]{revtex4-1}

\usepackage{graphicx}
\usepackage{dcolumn}
\usepackage{amsmath}
\usepackage{amsfonts}
\usepackage{amssymb}
\usepackage{bm}
\usepackage{bbm}
\usepackage{mathrsfs}
\usepackage{textcomp}
\usepackage{hyperref}
\usepackage{color}

\newcommand*{\op}[1]{\fontdimen12\textfont3=2pt\fontdimen12\scriptfont3=1.4pt\!\null\mathop{\vphantom{#1}\smash{#1}}\limits_{\sim}\null\!}
\newcommand*{\opDag}[1]{\op{#1}^{\dagger}}
\newcommand*{\opVec}[1]{\op{\bm{#1}}}

\newcommand*{\calOp}[1]{\op{\mathcal{#1}}}

\begin{document}


\title{Numerical Renormalization Group calculations of the magnetization of isotropic and anisotropic Kondo impurities}

\author{Martin \surname{H\"ock}}
\email{hoeck@physik.uni-bielefeld.de}
\author{J\"urgen \surname{Schnack}}
\email{jschnack@physik.uni-bielefeld.de}
\affiliation{Fakult\"at f\"ur Physik, Universit\"at Bielefeld, Postfach 100131, D-33501 Bielefeld, Germany}

\date{\today}

\begin{abstract}
We study a Kondo impurity model with additional uniaxial anisotropy $D$ in a non-zero magnetic field $B$ using the Numerical Renormalization Group (NRG). The ratio $g_e/g_S$ of electron and
impurity g-factor is regarded as a free parameter and, in particular, the special cases of a ``local'' ($g_e=0$) and ``bulk'' ($g_e=g_S$) field are considered. For a bulk field, the relationship between
the impurity magnetization $\mathcal{M}$ and the impurity contribution to the magnetization $M_{\text{imp}}$ is investigated. Furthermore, we study how the value of $g_e$ affects the impurity
magnetization curves. In case of an isotropic impurity with $g_e=g_S$, it is demonstrated that at zero temperature $\mathcal{M}(B)$, unlike $M_{\text{imp}}(B)$, does \emph{not} display universal
behavior. With additional ``easy axis'' anisotropy, the impurity magnetization is well described by a shifted and rescaled Brillouin function on energy scales that are small compared to $|D|$. In case
of ``hard axis'' anisotropy, the magnetization curves can feature steps which are due to field-induced pseudo-spin-1/2 Kondo effects. For large anisotropy and a local field, these screening effects
are described by an anisotropic spin-1/2 Kondo model with an additional scattering term that is spin-dependent (in contrast to ordinary potential scattering). Our study is motivated by the question
how the magnetic properties of a deposited magnetic molecule are modified by the interaction with a non-magnetic metallic surface.
\end{abstract}

\pacs{73.20.Hb, 75.30.Cr, 75.30.Gw, 75.50.Xx, 75.60.Ej}

\maketitle


\section{\label{sec:intro}Introduction}

Magnetic molecules offer the prospect of encoding and storing information in their magnetic state. The latter point applies, in particular, to bistable molecules such as single molecule magnets (SMMs).
The possibility to store, e.g., one bit of information in the state of a single molecule would constitute an enormous degree of miniaturization and could lead to data storage technologies with significantly
increased areal density. \cite{Rogez2009} However, to make a (potentially elusive) technological application feasible, the molecules need to be individually addressable so that their magnetic state can
be probed and manipulated on a molecule-by-molecule basis. In the last years, there has been an increasing interest in the question whether this functionality can be achieved by a controlled deposition
of magnetic molecules on suitable substrates. \cite{Gatteschi2009,Rogez2009,Cornia2011,Domingo2012} While such an approach might solve the problem of addressability, it can introduce new
complications due to interactions between the molecules and the surface. Depending on details such as the molecule's ligands, the presence of an additional decoupling layer, and, of course, the
characteristics of the surface, the interaction with the substrate might alter the magnetic properties of the molecule in an important (and possibly adverse) way. Thus, even if the magnetic response of
the isolated molecule is well understood (e.g., through a description by a suitable spin model \cite{GatteschiBook}), its magnetic properties in contact with the surface have to be reinvestigated.

In this article, we study a single-channel Kondo impurity model with non-zero magnetic field and additional uniaxial anisotropy $D (\op{S}^z)^2$ for the impurity spin operator $\opVec{S}$. Such
an anisotropy term (along with transverse anisotropy $E[(\op{S}^x)^2 - (\op{S}^y)^2]$) is a common part of a pure spin model for the description of isolated magnetic molecules (in particular, for
representing SMMs). \cite{GatteschiBook} The quantum impurity model is intended to serve as a minimal representation of an anisotropic magnetic molecule on a non-magnetic metallic substrate
and, with transverse anisotropy $E$, has already been used to describe SMMs interacting with metallic electrodes. \cite{Romeike2006a,Romeike2006b,*Romeike2011,Roosen2008,*Roosen2010}
Furthermore, it has been found that the above uniaxial and transverse anisotropy terms are also appropriate to model the surface-induced anisotropy of a single magnetic atom on a metallic
substrate with a decoupling layer. \cite{Hirjibehedin2007,Otte2008,Brune2009} To investigate how the interaction with the electrons affects the magnetic properties of the impurity, we carry out
Numerical Renormalization Group\cite{Wilson1975,Krishnamurthy1980a,Bulla2008} (NRG) calculations and focus on the magnetic field dependence of the impurity magnetization.

Regarding the experimental situation, the magnetic moment of deposited molecules (or atoms)\cite{Brune2009} can be measured using methods such as X-ray magnetic circular dichroism (XMCD).
\cite{Cornia2011,Mannini2009b,Mannini2009,Corradini2009,Stepanow2010,Mannini2010,Biagi2010,Gonidec2011,LodiRizzini2011,Corradini2012} XMCD is an element-specific technique of high
sensitivity based on the absorption of circularly polarized X-rays and can be used to obtain an ensemble-averaged result for the magnetic field dependent molecule magnetization. In principle, it is
also possible to extract information about different contributions to the observed magnetic moment (such as the orbital and spin contribution) from the XMCD data using, e.g., sum rules.
\cite{Cornia2011,Corradini2009,Stepanow2010,Biagi2010,Corradini2012} In the last years, magnetization curves of magnetic atoms on non-magnetic metallic surfaces could also be
recorded using spin-polarized scanning tunneling spectroscopy (SP-STS). \cite{Meier2008,Brune2009,Zhou2010,Wiebe2011,Khajetoorians2011,Khajetoorians2012} In contrast to XMCD, this
method provides a time-average of the field-dependent magnetic moment of a single atom. It has been demonstrated that SP-STS can also be applied to (suitable) deposited magnetic molecules.
\cite{Iacovita2008, Heinrich2010, Brede2012}

The static magnetization of Kondo impurity models (including related models such as the single-impurity Anderson model) has been investigated by a number of techniques. Among these are
Green's-function methods, \cite{Bloomfield1970,Poo1975a,*Poo1975b} the Bethe Ansatz, \cite{Fateev1981b, Andrei1981, Fateev1981a, Furuya1982, Rajan1982, Tsvelick1983, Andrei1983,
Schlottmann1983, Lowenstein1984, Tsvelick1985, Sacramento1989, Takegahara1990} and NRG \cite{Hewson2006, Wright2011, Zitko2011b} (including density matrix based extensions). By
now, there are also several studies of the time dependence of the magnetization in non-equilibrium situations (e.g., after a quantum quench or with a non-zero voltage bias). \cite{Hackl2009b,
Fritsch2010, Schiro2010, Pletyukhov2010, Heyl2010} In particular, non-equilibrium spin dynamics of impurity models can be investigated by using a generalization of NRG called time-dependent
NRG (TD-NRG). \cite{Anders2005,*Anders2006,Roosen2008,*Roosen2010,Hackl2009}

The present article extends existing NRG results for the Kondo model with uniaxial anisotropy \cite{Zitko2008b} to the case of non-zero magnetic field. The system with non-zero field (with a focus
on the properties of spectral functions) has been previously studied in Refs. \onlinecite{Zitko2009} and \onlinecite{Zitko2010c}. Furthermore, magnetization curves for isotropic and anisotropic
Kondo impurities have been calculated in Ref. \onlinecite{Zitko2011b}. We would like to stress, however, that our investigation places emphasis on different aspects of the problem and is thus
complementary to Ref. \onlinecite{Zitko2011b}.

The remainder of this article is organized as follows. In Sec. \ref{sec:model}, the quantum impurity model is introduced and transformed to a representation that is suitable for further numerical
treatment. Sec. \ref{sec:methodObs} provides information about our use of the NRG method and contains definitions of the considered observables. In Sec. \ref{sec:isotropic}, we study the
relationship between the impurity magnetization and the impurity contribution to the magnetization for an isotropic system ($D=0$). After an investigation of the Kondo model with additional
``easy axis'' anisotropy ($D<0$) in Sec. \ref{sec:easyAxis}, the case of ``hard axis'' anisotropy ($D>0$) is considered in Sec. \ref{sec:hardAxis}. There, we also clarify how a non-zero magnetic
coupling of the conduction electrons affects the impurity magnetization. Furthermore, an effective model is derived in order to describe the field-induced pseudo-spin-1/2 Kondo effects that are
observed in the magnetization curves for large $D$. We conclude this article with a summary of the results in Sec. \ref{sec:summary} and a brief description of the technical details of an NRG
calculation with non-zero magnetic coupling of the conduction electrons in App. \ref{app:NRGWithElectronZeeman}.

\section{\label{sec:model}Model}

\subsection{\label{subsec:H}Hamiltonian}

In this work, we study a Hamilton operator $\op{H}$ consisting of three parts:

\begin{equation}
	\op{H} = \op{H}_{\text{electrons}} + \op{H}_{\text{coupling}} + \op{H}_{\text{impurity}} \; .
	\label{eq:HTotal}
\end{equation}

The first term, $\op{H}_{\text{electrons}}$, represents non-interacting tight-binding electrons whose hopping between different sites $i$ and $j$ of a periodic lattice with $L$ sites is described by the
corresponding hopping parameter $t_{ij}$:

\begin{equation}
	\op{H}_{\text{electrons}} = \sum_{i \neq j, \, \sigma}{t_{ij} \opDag{d}_{i\sigma} \op{d}_{j\sigma}} + g_e \mu_B B \calOp{S}^z  \; .
	\label{eq:HElectronsReal}
\end{equation}

\noindent Here, $\op{d}^{(\dagger)}_{{i \sigma}}$ is a destruction (creation) operator for an electron with spin projection $\sigma = \pm 1/2 \; \widehat{=} \uparrow / \downarrow$ at lattice site $i$. The effect of
an external magnetic field $B$ is taken into account by a Zeeman term with electron g-factor $g_e$, Bohr magneton $\mu_B$, and the $z$-component of the total spin of the electrons
$\calOp{S}^z = \frac{1}{2} \sum_{i}{(\op{n}_{i \uparrow} - \op{n}_{i \downarrow})}$ with $\op{n}_{i \sigma} = \opDag{d}_{i \sigma} \op{d}_{i \sigma}$. Using a discrete Fourier transformation,
$\opDag{c}_{\bm{k} \sigma} = (1/\sqrt{L}) \sum_{j}{e^{i \bm{k} \cdot \bm{R}_j} \opDag{d}_{j \sigma}}$, Hamiltonian (\ref{eq:HElectronsReal}) can be equivalently written in the more common form

\begin{equation}
	\op{H}_{\text{electrons}} = \sum_{\bm{k},\sigma}{\underbrace{(\varepsilon_{\bm{k}} + \sigma g_e \mu_B B)}_{= \, \varepsilon_{\bm{k}\sigma}(B)} \opDag{c}_{\bm{k}\sigma} \op{c}_{\bm{k}\sigma}} \; ,
\end{equation}

\noindent with a dispersion relation $\varepsilon_{\bm{k}\sigma}(B)$, assigning an energy $\varepsilon$ to a wavevector $\bm{k}$, that now depends on spin projection and magnetic field. In general,
the spin-independent dispersion relation $\varepsilon_{\bm{k}}$ is anisotropic in $\bm{k}$-space.

For the interaction term in Eq. (\ref{eq:HTotal}), we use a standard isotropic Kondo coupling,

\begin{equation}
	\op{H}_{\text{coupling}} = J \opVec{S} \cdot \opVec{s}_0 \; ,
	\label{eq:KondoCoupling}
\end{equation}

\noindent and assume that the impurity spin $\opVec{S}$ couples antiferromagnetically ($J>0$) to the electronic spin at the origin, which is given by
$\opVec{s}_0 = (1/2L) \sum_{\bm{k}, \bm{k}', \mu,\nu}{\opDag{c}_{\bm{k} \mu} \bm{\sigma}_{\mu \nu} \op{c}_{\bm{k}' \nu}}$ with the vector of Pauli matrices $\bm{\sigma}$.

Finally, the impurity part of Hamiltonian (\ref{eq:HTotal}) represents a localized spin with quantum number $S$ which couples to the external magnetic field with g-factor $g_S$ and possesses an
additional uniaxial anisotropy $D$:

\begin{equation}
	\op{H}_{\text{impurity}} = D ( \op{S}^z )^2 + g_S \mu_B B \op{S}^z \; .
	\label{eq:HImp}
\end{equation}

\noindent With the chosen convention, the impurity spin has an ``easy axis'' for $D<0$ and a ``hard axis'' or an ``easy plane'' for $D>0$. A further transverse anisotropy $E[(\op{S}^x)^2 - (\op{S}^y)^2]$
is \emph{not} considered in this article. $\op{H}_{\text{impurity}}$ can be seen as a minimal representation of a magnetic molecule with a single magnetic center or as a ``giant spin approximation''
for an SMM. \cite{GatteschiBook,BenciniBook}

Hamiltonian (\ref{eq:HTotal}) corresponds to an anisotropic single-channel Kondo impurity model in an external magnetic field. The special choices $g_e=0$ and $g_e=g_S$ for the electron g-factor
are referred to as a ``local'' and ``bulk'' magnetic field, respectively. Regarding the modeling of a deposited magnetic molecule, it has to be emphasized that the Hamiltonian suffers from a number of
simplifications. For example, there is no orbital contribution to the magnetism, and no charge fluctuations between molecule and surface are possible. In this article, we only consider the effect of the
Kondo coupling on the magnetic properties of the impurity spin.

\subsection{\label{subsec:energyRep}Transformation to an energy representation}

In order to treat Hamiltonian (\ref{eq:HTotal}) using NRG, $\op{H}_{\text{electrons}}$ and $\op{H}_{\text{coupling}}$ are expressed via a continuous energy representation for the electronic degrees
of freedom. To this end, we first take a standard continuum limit in $\bm{k}$-space (i.e., we consider a lattice of dimension $d$ with $L \gg 1$). \cite{Krishnamurthy1980a} By adapting the corresponding
expression for the two-impurity Kondo model from Ref. \onlinecite{Affleck1995} to the single-impurity case (see also Ref. \onlinecite{Schiller2008}), we then define those states with energy
$\varepsilon$ to which the localized spin directly couples:

\begin{equation}
	\op{a}_{\varepsilon \mu} = \frac{1}{\sqrt{(2 \pi)^d \rho(\varepsilon - \mu h)}} \int{\mathrm{d} \bm{k} \, \delta(\varepsilon - \varepsilon_{\mu}(\bm{k}, B)) \, \op{c}_{\bm{k} \mu}} \; ,
	\label{eq:energyState}
\end{equation}

\noindent where we have introduced the abbreviation $h = g_e \mu_B B$ and the normalized density of states (DOS) per spin projection and lattice site
$\rho(\varepsilon) = (1/L) \sum_{\bm{k}}{\delta(\varepsilon - \varepsilon_{\bm{k}})}$. Denoting the half-width of the conduction band by $W$, the allowed
energies $\varepsilon$ for spin projection $\mu$ span the interval $[-W + \mu h, \, W + \mu h]$. The new operator $\op{a}_{\varepsilon \mu}$ is properly
normalized because of the pre-factor involving the DOS.

If we are only interested in impurity properties, then all other electronic states different from those defined in Eq. (\ref{eq:energyState}) can be safely dropped
without introducing any approximation. \cite{Krishnamurthy1980a} This leads to the desired continuous energy representation of Hamiltonian (\ref{eq:HTotal}): 

\begin{eqnarray}
	\op{H} & \rightarrow & \sum_{\mu}{\int_{-W+\mu h}^{W+\mu h}{\mathrm{d} \varepsilon \, \varepsilon \, \opDag{a}_{\varepsilon \mu} \op{a}_{\varepsilon \mu}}} \nonumber \\
	& + & J \opVec{S} \cdot \sum_{\mu, \nu}{\left( \int_{-W+\mu h}^{W+\mu h}{\mathrm{d} \varepsilon \, \sqrt{\rho(\varepsilon - \mu h)} \, \opDag{a}_{\varepsilon \mu}} \right) \frac{\bm{\sigma}_{\mu \nu}}{2}} \times \nonumber \\
	& & \left( \int_{-W+\nu h}^{W+\nu h}{\mathrm{d} \varepsilon' \, \sqrt{\rho(\varepsilon' - \nu h)} \, \op{a}_{\varepsilon' \nu}} \right) + \op{H}_{\text{impurity}} \, .
	\label{eq:energyRep}
\end{eqnarray}

\noindent For $h = 0$, i.e., for $B=0$ or $g_e = 0$, Eq. (\ref{eq:energyRep}) reduces to the well-known expression for the energy representation of the Kondo model. \cite{Costi1999b}
In the following, we consider the case of a constant DOS: $\rho(\varepsilon) = 1/2W = \rho$.

\section{\label{sec:methodObs}Method and observables}

\subsection{\label{subsec:method}Method: NRG}

Approximate eigenvalues and eigenvectors of Hamiltonian (\ref{eq:energyRep}) for the calculation of impurity properties can be obtained using the Numerical Renormalization Group \cite{Wilson1975,
Krishnamurthy1980a, Bulla2008} (NRG). However, the procedure leading to the parameters of the Wilson chain has to be slightly modified if $h \neq 0$ (see App. \ref{app:NRGWithElectronZeeman} for
a brief discussion of the required changes).

A non-zero magnetic field breaks the full SU(2)-symmetry in spin space of Hamiltonian (\ref{eq:HTotal}). For this reason, we label eigenstates of $\op{H}$ only with the charge quantum number $Q$ and the
magnetic quantum number $S^z_{\text{total}}$ of the $z$-component of the total spin. Except for one example in Sec. \ref{sec:hardAxis}, all NRG calculations are carried out using the discretization scheme
proposed by \v{Z}itko and Pruschke \cite{Zitko2009a,Zitko2009b} with averaging over 4 $z$-values that are equidistantly spaced on the interval $(0,1]$. Observables are computed using only states that are
kept after truncation and results are averaged over even and odd sites of the Wilson chain according to the prescription of Ref. \onlinecite{Bulla2008}. We use a discretization parameter
$\Lambda = 3$, a dimensionless inverse temperature $\bar{\beta} = 0.7$, and a fixed number of kept states of the order of $N=5000$ to achieve convergence for all considered observables within the
resolution of the presented plots. Nevertheless, at $\Lambda>1$ there might still be slight systematic deviations for non-zero temperature, which can for example be demonstrated by setting $J=0$ and
comparing the NRG results with the analytical solution for a free spin. It is necessary to perform a separate NRG calculation for each value of the magnetic field. If curves are shown in a plot, they are thus the
result of a spline interpolation through the numerically obtained data points.

\subsection{\label{subsec:obs}Observables}

In our calculations we focus on the impurity magnetization which is defined as the expectation value of the impurity magnetization operator with respect to the eigenstates of the total Hamiltonian (\ref{eq:HTotal}):

\begin{equation}
	\mathcal{M}(T,B) = - \left \langle \frac{\partial \op{H}_{\text{impurity}}}{\partial B} \right \rangle_{\text{total}} = - g_S \mu_B \langle \op{S}^z \rangle_{\text{total}} \; .
	\label{eq:impMagnetization}
\end{equation}

Furthermore, we consider the impurity contribution to the entropy, magnetization, and magnetic susceptibility. The impurity contribution to some quantity $\mathcal{O}$ is defined in the usual way: \cite{Bulla2008}

\begin{equation}
	\mathcal{O}_{\text{imp}} = \mathcal{O}_{\text{total}}^{\text{with impurity}} - \mathcal{O}_{\text{total}}^{\text{w/o impurity}} \; .
	\label{eq:impContribution}
\end{equation}

\noindent The observable $\mathcal{O}_{\text{total}}^{\text{w/o impurity}}$ for the system without impurity is also calculated using NRG by removing the impurity part from the Wilson chain. For the entropy
$S(T,B)$, the magnetization $M(T,B)$, and the susceptibility $\chi(T,B)$, we use the standard definitions $S(T,B) = - \partial \Omega(T,B)/\partial T$, $M(T,B) = - \partial \Omega(T,B)/ \partial B$,
and $\chi(T,B) = \partial M(T,B)/ \partial B$, with $\Omega(T,B)$ being the grand-canonical potential. If the electron g-factor is zero, we have the special case $\mathcal{M}(T,B) = M_{\text{imp}}(T,B)$.

In the grand-canonical calculations the chemical potential is assumed to be zero. For a symmetric DOS, $\rho(\varepsilon) = \rho(-\varepsilon)$, the free electron band is thus on average half-filled for
arbitrary magnetic field and temperature.

\section{\label{sec:isotropic}Isotropic impurities}

Let us first consider the case of an isotropic impurity with $D = 0$ in Hamiltonian (\ref{eq:HImp}) and study the impurity contribution to the magnetization $M_{\text{imp}}$ and the impurity magnetization
$\mathcal{M}$, both as function of temperature and magnetic field. For the moment, we are only concerned with the special case of equal g-factors of impurity and electrons (corresponding to the case of
a bulk magnetic field). Recalling the motivation given in the introduction, $\mathcal{M}$ as the expectation value of the impurity magnetization operator should be the observable that is more closely related
to experimental magnetization data obtained by methods such as XMCD. Note that $\mathcal{M}$ and $M_{\text{imp}}$ become equivalent if impurity and electrons decouple (which happens for $J \rightarrow 0$
or $T \rightarrow \infty$).

\subsection{\label{subsec:M_vs_B-isotrop}Field dependence of the magnetization}

In case of the isotropic Kondo model with $g_e = g_S$ and arbitrary impurity spin $S$, the Bethe Ansatz (BA) allows for the derivation of a closed expression for the impurity contribution to the magnetization
at zero temperature. \cite{Fateev1981b, Andrei1981, Fateev1981a, Furuya1982, Tsvelick1983, Andrei1983} $M_{\text{imp}}$ is known to display universal behavior in the so-called scaling regime, in which all
relevant energy scales are small compared to the energy cutoff (or the finite bandwidth). \cite{Andrei1983} Bare parameters of the model can then be absorbed into an energy scale $k_B T_H$ so that the
field-dependence of $M_{\text{imp}}$ at $T=0$ is described by a universal function $f(x)$, with $x$ being the rescaled magnetic field: $x = g_S \mu_B B / k_B T_H$. \cite{Andrei1983} $T_H$ is related to the
Kondo temperature $T_K$ by a constant factor: \cite{Tsvelick1983}

\begin{equation}
	T_H = \sqrt{\frac{2 \pi}{e}} \; T_K \; .
	\label{eq:THandTK} 
\end{equation}

\begin{figure}
\centering{\includegraphics[width=1.0 \linewidth]{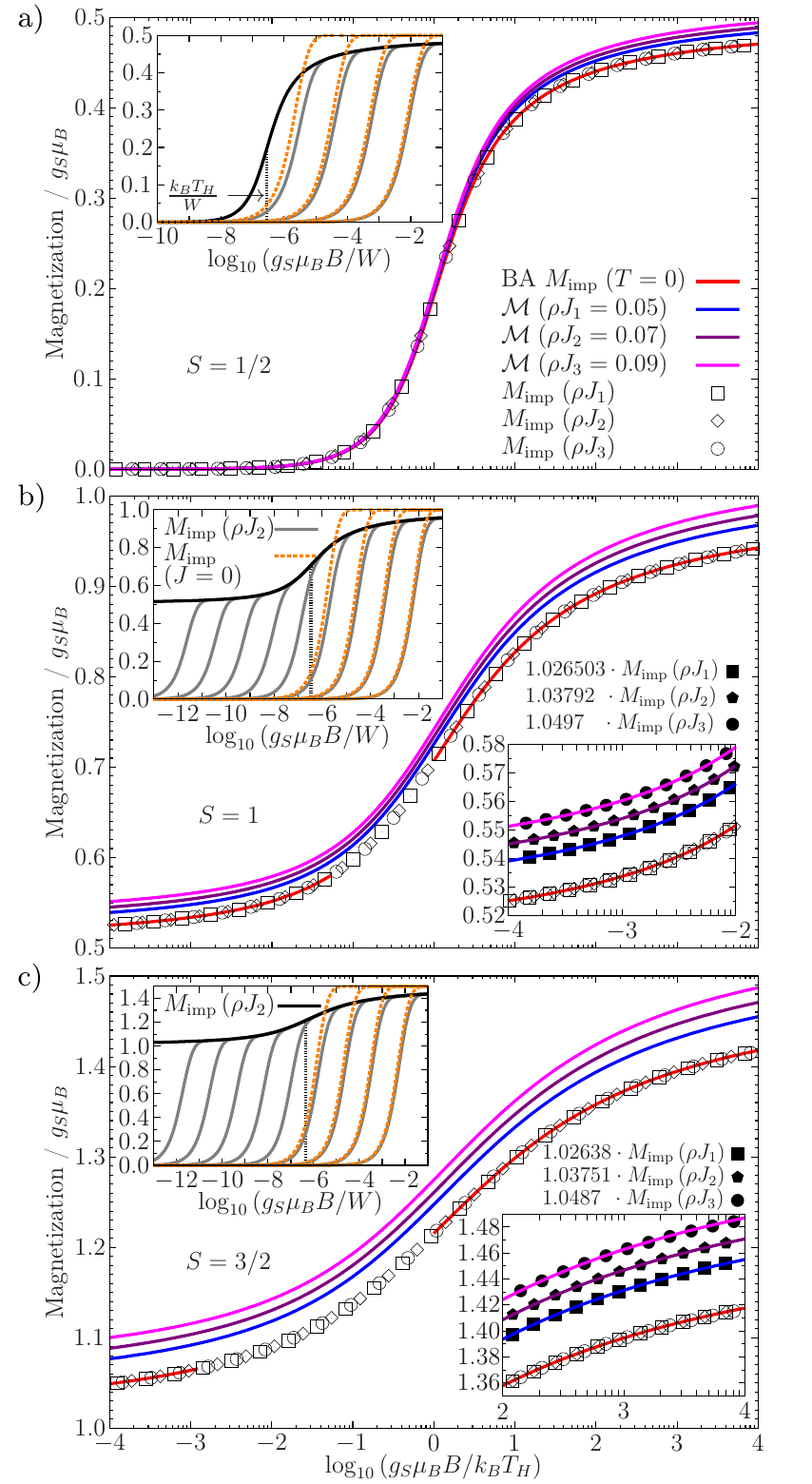}}
%
%
%
%
\caption{\label{fig:M_vs_B-isotrop}(Color online) \textbf{Main plots:} Impurity contribution to the magnetization $M_{\text{imp}}$ and impurity magnetization $\mathcal{M}$ as function of magnetic field for
$g_e=g_S$, three different couplings $\rho J$, and for impurity spin a) $S=1/2$, b) $S=1$, and c) $S=3/2$. The temperature is $k_B T/W \approx 1.54 \cdot 10^{-15} \approx 0$ and the field is rescaled
using $k_B T_H$. For the shown $\mathcal{M}$-curves the coupling $\rho J$ always increases from bottom to top. The universal BA solution at $T=0$ has only been calculated for $g_S\mu_B B \geq k_B T_H$
and $g_S \mu_B B \ll k_B T_H$ in plots b) and c). \textbf{Upper left insets} show NRG results for $M_{\text{imp}}$ at $T \approx 0$ (black lines) and finite temperature (solid gray lines) as a function of magnetic
field, now expressed in units of $W$. $M_{\text{imp}}$ for $J=0$ is also computed using NRG and resembles the magnetization of the free spin. Thermal energies increase from left to right and range from
$1.79 \cdot 10^{-6}W$ (plot a)) or $1.95 \cdot 10^{-12}W$ (plots b) and c)) to $6.79 \cdot 10^{-3}W$. \textbf{Lower right insets} show a close-up of the magnetization curves for b) low fields and c) high fields
along with data points for $M_{\text{imp}}$ that are multiplied by a constant $\gtrsim 1$.}
\end{figure}

In Fig. \ref{fig:M_vs_B-isotrop} we plot the Bethe Ansatz solution for $M_{\text{imp}}$ from Ref. \onlinecite{Tsvelick1983} for three different impurity spins $S=1/2,1$, and $3/2$. \footnote{Note that for higher
spin $S \geq1$ part of the curve for $g_S \mu_B B \lesssim k_B T_H$ is missing. The reason for this is that the real-valued integral representation for $M_{\text{imp}}$ from Ref. \onlinecite{Tsvelick1983} is
only asymptotically correct for $g_S \mu_B B < k_B T_H$, i.e., it becomes exact only for $g_S \mu_B B \ll k_B T_H$.} $f(x)$ is a strictly monotonically increasing function of $x$ and approaches the saturation
magnetization of a free spin, $g_S \mu_B S$, for $x \rightarrow \infty$ with slowly decaying logarithmic corrections. \cite{Tsvelick1983,Andrei1983,HewsonBook} The behavior in the limit $x \rightarrow 0$
depends on the value of $S$: In case of $S=1/2$, $f(x) \propto x$ for small $x$, whereas for $S \geq 1$ the function $f(x)$ goes to the saturation magnetization of a reduced spin with $S-1/2$, again with
logarithmic corrections. \cite{Tsvelick1983,Andrei1983,HewsonBook} This low-field behavior mirrors the Kondo screening which, for vanishing magnetic field, reduces the impurity spin $S$ to a residual spin
$S-1/2$ in the limit $T/T_K \ll 1$. \cite{Mattis1967,Cragg1979a} The magnetic properties of the impurity are furthermore markedly different from that of a free spin as the magnetization of a free spin at $T=0$
saturates for any non-zero magnetic field.

Using NRG, we have calculated $M_{\text{imp}}$ for several values of the coupling strength $\rho J$ and have fitted the obtained curves to the respective universal BA curve by employing $T_H$ as a fit
parameter (see Fig. \ref{fig:M_vs_B-isotrop}). The nice agreement with the Bethe Ansatz solution demonstrates the universal behavior that $M_{\text{imp}}$ displays for small $\rho J$ and allows us to reliably
determine the value of $T_H$ even for impurity spin $S \geq 1$. Note that for very large magnetic fields, i.e., for $g_S \mu_B B \lesssim W$, we leave the scaling regime and $M_{\text{imp}}$, as calculated by
NRG, starts to drop below the universal BA curve (this is not shown in Fig. \ref{fig:M_vs_B-isotrop}). The determined approximate values of $k_B T_H/W$ are found in Table \ref{tab:T_H-isotropic}. In the cutoff
scheme used in the BA solution of the Kondo model with arbitrary impurity spin, the dynamically generated low-energy scale $T_0$, whose ratio to $T_H$ is a universal number, depends on the value of $S$
(before taking the scaling limit). \cite{Andrei1981, Furuya1982, Rajan1982, Andrei1983} Here, we find that the fitted values of $T_H$ increase with the impurity spin for fixed coupling strength and, furthermore,
that the relative deviation between the results for different $S$ decreases when $\rho J$ is reduced. However, even for the smallest considered coupling strength ($\rho J = 0.05$), the values of $T_H$ for $S=1/2$
and $S=3/2$ still deviate by about 44~\%. According to Eq. (\ref{eq:THandTK}), the values of $T_H$ for $S=1/2$ reported in Table \ref{tab:T_H-isotropic} correspond to the following Kondo temperatures:
$k_B T_K / W \approx 4.79 \cdot 10^{-10} \; (\rho J=0.05)$, $1.80 \cdot 10^{-7} \; (0.07)$, and $5.08 \cdot 10^{-6} \; (0.09)$. These results can be compared with the standard estimate for $T_K$, \cite{Wilson1975,
Krishnamurthy1980a}

\begin{equation}
	k_B T_K / W \approx \sqrt{\rho J} \exp{(-1/\rho J)} \; ,
	\label{eq:TKEstimate}
\end{equation}

\noindent which is valid for small coupling and gives $k_B T_K / W \approx 4.61 \cdot 10^{-10} \; (\rho J=0.05)$, $1.65 \cdot 10^{-7} \; (0.07)$, and $4.48 \cdot 10^{-6} \; (0.09)$. As a further check, we have
determined the Kondo temperature for $S=1/2$ and $\rho J = 0.07$ by fitting the BA solution for the impurity contribution to the susceptibility from Ref. \onlinecite{Tsvelick1983} and the impurity contribution
to the entropy from Ref. \onlinecite{Desgranges1982}. This gives a value of $k_B T_K/W \approx 1.79 \cdot 10^{-7}$, which is quite similar to the one following from Table \ref{tab:T_H-isotropic}.

\begin{table*}
\caption{\label{tab:T_H-isotropic}Approximate values of $k_B T_H/W$ as used in Fig. \ref{fig:M_vs_B-isotrop}, obtained by fitting the universal Bethe Ansatz solution for $M_{\text{imp}}$, and
proportionality factors $\alpha(\rho J)$ relating $\mathcal{M}$ and $M_{\text{imp}}$ according to $\mathcal{M} = \alpha M_{\text{imp}}$. The results for $\alpha$ have been averaged over magnetic
fields $g_S \mu_B B / W \in [10^{-13}, 10^{-1}]$ for $k_BT/W \approx 1.54 \cdot 10^{-15} \approx 0$. Numbers in parentheses give the corresponding standard deviation for the last decimal place.
For Zeeman energies close to the band edge (i.e., $g_S \mu_B B \lesssim W$), which are not considered for the average, $\alpha$ noticeably decreases (increases) for $S=1/2$
($S=1, 3/2$).}
\begin{ruledtabular}
\begin{tabular}{cllllll}
& \multicolumn{2}{c}{$S=1/2$} & \multicolumn{2}{c}{$S=1$} & \multicolumn{2}{c}{$S=3/2$} \\
$\rho J$ & $k_B T_H/W$ & \multicolumn{1}{c}{$\alpha$} & $k_B T_H/W$ & \multicolumn{1}{c}{$\alpha$} & $k_B T_H/W$ & \multicolumn{1}{c}{$\alpha$} \\
\colrule
0.05 & $7.29 \cdot 10^{-10}$ & 1.02659(1) & $8.49 \cdot 10^{-10}$ & 1.026503(7) & $1.05 \cdot 10^{-9}$ & 1.02638(2) \\
0.07 & $2.74 \cdot 10^{-7}$   & 1.03822(2) & $3.39 \cdot 10^{-7}$   & 1.03792(3)    & $4.55 \cdot 10^{-7}$ & 1.03751(6) \\
0.09 & $7.72 \cdot 10^{-6}$   & 1.05048(3) & $1.02 \cdot 10^{-5}$   & 1.04970(8)    & $1.51 \cdot 10^{-5}$ & 1.0487(2) \\
\end{tabular}
\end{ruledtabular}
\end{table*}

The upper left insets of Fig. \ref{fig:M_vs_B-isotrop} show finite temperature NRG results for $M_{\text{imp}}$ with a coupling strength $\rho J=0.07$. While the Bethe Ansatz provides a closed expression for
$M_{\text{imp}}$ at zero temperature, a calculation for finite temperature leads to so-called thermodynamic BA equations that, at least in general, have to be solved numerically. \cite{Rajan1982, Okiji1983}
Hence, finite temperature results for the magnetization are not easily available. As a reference point, we replot the zero temperature magnetization curve that crosses over to the strong coupling regime in the
vicinity of $g_S \mu_B B \approx k_B T_H$. As long as the thermal energy is small compared to the Zeeman energy, the magnetization always closely follows the zero temperature curve. On the other hand,
if the thermal energy is not negligibly small compared to the Zeeman energy, we have to distinguish between complete screening and underscreening of the impurity spin. For $S \geq 1$, finite temperature
is always important as it also affects the residual spin. On the energy scale $g_S \mu_B B \approx k_B T$ there is a swift drop of $M_{\text{imp}}$ that is eventually followed by a linear decay for small fields
$g_S \mu_B B \ll k_B T$. In the special case $S=1/2$, however, finite temperature has little effect if $T \ll T_K$ and the magnetization already displays a linear dependence on the magnetic field for
$g_S \mu_B B \approx k_B T$ due to the Kondo screening. In the upper left insets of Fig. \ref{fig:M_vs_B-isotrop} we also compare the results for $M_{\text{imp}}$ with NRG calculations for vanishing coupling
$J = 0$. This comparison is meant to illustrate the influence of a non-zero value of $J$. \footnote{It might seem more natural to compare with the easily obtainable magnetization of a free spin. However, we
have observed that at zero coupling NRG is apparently not able to fully reproduce the finite temperature behavior of a free spin for the chosen value of $\Lambda$. For this reason, we rather use NRG results
with $J=0$ for the comparison} At high temperatures (compared to the Kondo temperature) the impurity is progressively decoupled from the electronic system and its magnetization will hence resemble the
result for $J=0$ more closely. However, note that the impurity only becomes asymptotically free for high temperatures.

In addition to the impurity contribution to the magnetization $M_{\text{imp}}$, we also plot the impurity magnetization $\mathcal{M}$ in Fig. \ref{fig:M_vs_B-isotrop} for the same values of the coupling $\rho J$
and negligible temperature. The magnetic field is again rescaled by $k_B T_H$ using the values from Table \ref{tab:T_H-isotropic}. We find that $\mathcal{M}$ and $M_{\text{imp}}$ differ for all considered
magnetic fields with $\mathcal{M}$ being larger than $M_{\text{imp}}$. This means, in particular, that for large magnetic fields $\mathcal{M}$ comes closer to the saturation magnetization of a free spin than
$M_{\text{imp}}$ does and, furthermore, that the magnetization of the conduction electrons is reduced due to the interaction with the impurity spin. Upon decreasing $\rho J$ we observe that the impurity
magnetization becomes smaller and thus approaches the universal curve for $M_{\text{imp}}$. A comparison of the NRG results for $\mathcal{M}$ and $M_{\text{imp}}$ shows that both quantities are
proportional to each other, i.e., $\mathcal{M} = \alpha M_{\text{imp}}$ with a proportionality factor $\alpha > 1$ that depends on the coupling strength $\rho J$ (see Table \ref{tab:T_H-isotropic} for a list of the
computed values of $\alpha$). This proportionality is illustrated for the case of small magnetic fields (for $S=1$) and large magnetic fields (for $S=3/2$) in the lower right insets of Fig. \ref{fig:M_vs_B-isotrop}.
While the obtained values for $\alpha$ decrease with increasing impurity spin $S$, the values for different $S$ differ by less than 0.2~\% according to Table \ref{tab:T_H-isotropic}. Since impurity and electrons
progressively decouple at high temperatures, we expect $\alpha$ to be temperature dependent with $\alpha \rightarrow 1$ for $k_B T /W \gg 1$. The results presented in Fig. \ref{fig:M_vs_B-isotrop} show that
the magnetic field cannot be rescaled by $k_B T_H$ or any scale proportional to it so as to produce a universal curve for the impurity magnetization $\mathcal{M}$. 

To elucidate our findings we refer to one of the original Bethe Ansatz investigations of the Kondo model. \cite{Lowenstein1984} There it is found that, under the assumptions of a BA calculation (including
an arbitrarily large energy cutoff $\mathscr{D}$), $\mathcal{M} = M_{\text{imp}}$. To study the influence of the cutoff scheme, a comparison with perturbation theory is carried out showing that $\mathcal{M}$ has
leading corrections of the order $1/\ln(\mathscr{D})$, whereas the corrections of $M_{\text{imp}}$ vanish like $1/\mathscr{D}$ and thus much faster. \cite{Lowenstein1984} The regime in which all relevant energy
scales are negligible compared to the cutoff \emph{in a logarithmic sense}, e.g. $\ln(\mathscr{D} / g_S \mu_B B) \gg 1$, is termed ``extreme scaling limit''. \cite{Lowenstein1984}

With this background we reach the following interpretation of our NRG results for $\mathcal{M}$ and $M_{\text{imp}}$: For the chosen values of the coupling strength $\rho J$ the half-bandwidth $W$ (basically
serving as the unit of energy) can be regarded as very large compared to all relevant energy scales $\mathcal{E}$ so that corrections of order $\mathcal{E}/W$ can be expected to be small. It is for this reason
that we find nice agreement with the universal BA solution for $M_{\text{imp}}$. On the other hand, corrections of order $1/\ln(W/\mathcal{E})$ are not necessarily negligible for a finite value of $W$. This appears
to be an adequate explanation for our NRG results showing that $\mathcal{M} \neq M_{\text{imp}}$. Moreover, a decrease of $\rho J$ corresponds to an increase of the bandwidth and thus bandwidth-related
corrections should become smaller. Accordingly, $\mathcal{M}$ decreases for smaller coupling strength and approaches $M_{\text{imp}}$. These observations might also bear some importance for experimental
situations: While experimental parameters are certainly suitable to consider the scaling regime (in case the system exhibits universal behavior), it is less clear whether an experimental system can be placed in the
extreme scaling regime.



\subsection{\label{subsec:MSat_vs_J-isotrop}Dependence of the zero temperature magnetization on the coupling strength}

\begin{figure}
\centering{\includegraphics[width=1.0 \linewidth]{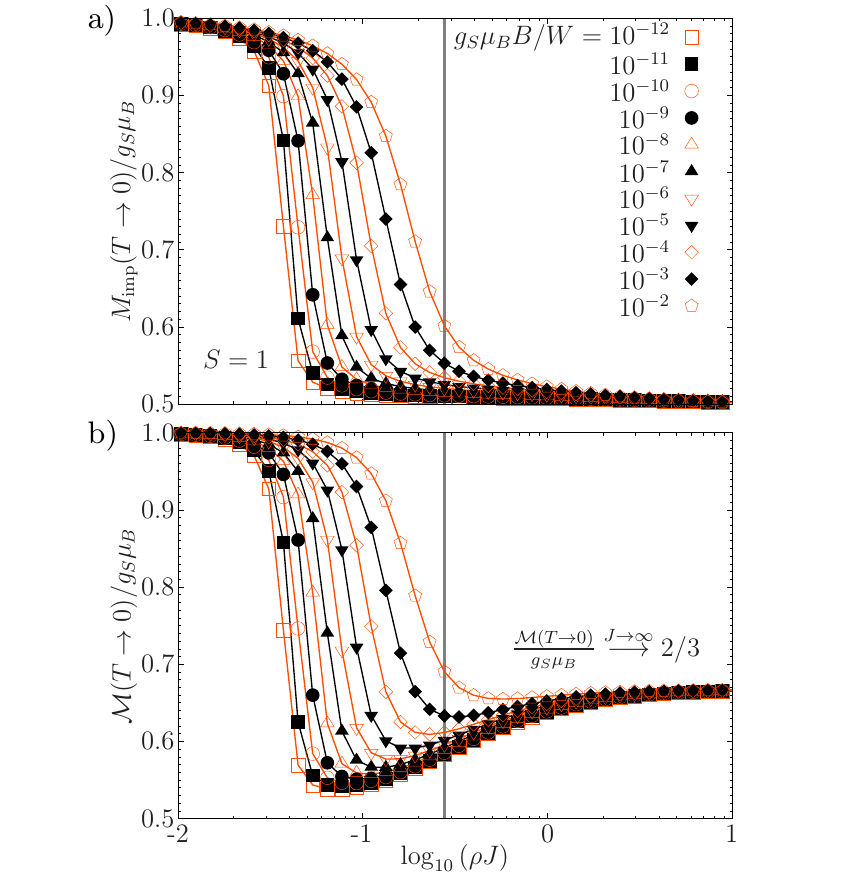}}
%
%
%
%
\caption{\label{fig:MSat_vs_J-isotrop-S1}(Color online) Saturation value in the limit $T \rightarrow 0$ of a) the impurity contribution to the magnetization $M_{\text{imp}}$ and b) the impurity magnetization
$\mathcal{M}$ as function of the coupling strength $\rho J$ for impurity spin $S=1$, $g_e=g_S$, and several magnetic field values. The vertical lines mark a coupling of $\rho J' \approx 0.276$ for which we have
checked by comparing with the Bethe Ansatz solution shown in Fig. \ref{fig:M_vs_B-isotrop} that $M_{\text{imp}}$ still exhibits universal behavior (a fit of the BA curve gives $k_B T_H/W \approx 8.85 \cdot 10^{-2}$).
Remaining lines are intended as a guide to the eye.}
\end{figure}

\begin{figure}
\centering{\includegraphics[width=1.0 \linewidth]{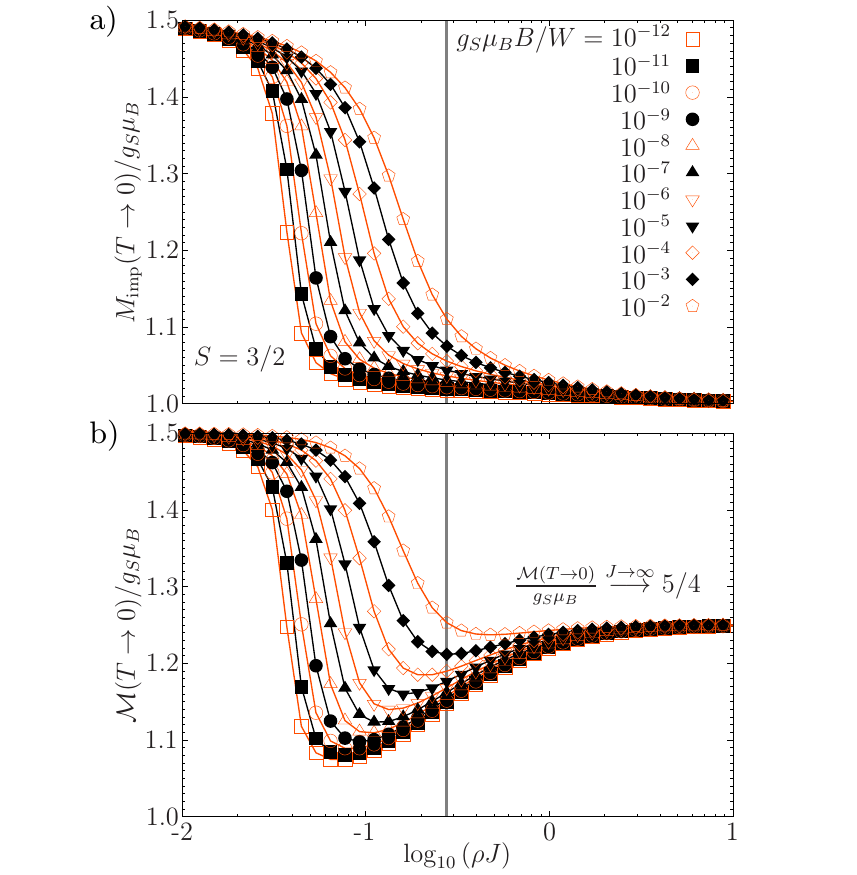}}
%
%
%
\caption{\label{fig:MSat_vs_J-isotrop-S3-2}(Color online) Saturation value in the limit $T \rightarrow 0$ of a) the impurity contribution to the magnetization and b) the impurity magnetization as function of the coupling
strength $\rho J$ as in Fig. \ref{fig:MSat_vs_J-isotrop-S1}, but here for impurity spin $S=3/2$. $M_{\text{imp}}$ again shows universal behavior for $\rho J' \approx 0.276$ and a fit of the universal BA curve now gives
$k_B T_H/W \approx 3.05 \cdot 10^{-1}$.}
\end{figure}

To further illustrate the difference between the impurity contribution to the magnetization and the impurity magnetization, we examine how both quantities depend on the coupling strength $\rho J$ for non-zero magnetic
field at zero temperature. As before, the case of equal g-factors for impurity and electrons is considered. NRG results for impurity spin $S=1$ and $S=3/2$ are shown in Fig. \ref{fig:MSat_vs_J-isotrop-S1} and Fig.
\ref{fig:MSat_vs_J-isotrop-S3-2}, respectively.

Let us begin the interpretation of the plots by considering the limiting cases $\rho J \rightarrow 0$ and $\rho J \rightarrow \infty$. For vanishing coupling strength $J \rightarrow 0$, impurity and electrons are decoupled
and thus both $M_{\text{imp}}$ and $\mathcal{M}$ correspond to the magnetization of a free spin which saturates for any non-zero value of the magnetic field at $T=0$. However, the behavior in the limit $J \rightarrow \infty$
differs for the two quantities, demonstrating that in general $M_{\text{imp}} \neq \mathcal{M}$. The values of $M_{\text{imp}}$ and $\mathcal{M}$ in this limit can be understood by considering a simplified model: In the
interaction term Eq. (\ref{eq:KondoCoupling}), the spin operator $\opVec{s}_0$ depends on the occupation of the lattice site that it is associated with. For very large values of $J$ it is energetically favorable that this lattice
site is singly occupied with a probability near one. We can then replace $\opVec{s}_0$ by a spin-1/2 operator $\opVec{s}$ in Eq. (\ref{eq:KondoCoupling}). Furthermore, all other terms in Hamiltonian (\ref{eq:HTotal})
that involve degrees of freedom different from the impurity and the lattice site to which it couples can be neglected. We are then left with a strongly coupled antiferromagnetic dimer in a magnetic field with Hamiltonian

\begin{equation}
\op{H}_{J \rightarrow \infty} = J \opVec{S} \cdot \opVec{s} + g_S \mu_B B (\op{S}^z + \op{s}^z) \; .
\label{eq:HEffectiveIso}
\end{equation}

To determine the limiting value of $\mathcal{M}/g_S \mu_B$ for $\rho J \rightarrow \infty$ at any positive magnetic field, we thus have to calculate the expectation value of $-\op{S}^z$ with respect to that eigenstate
of total spin which has the lowest value of $S_{\text{total}}$ and corresponding $z$-projection $M_{\text{total}} = - S_{\text{total}}$. Denoting eigenstates of the effective Hamiltonian (\ref{eq:HEffectiveIso}) as
$| S_{\text{total}}, M_{\text{total}} \rangle$, we find:

\begin{eqnarray}
\label{eq:JLimit1}
- \langle 0, 0 | \op{S}^z | 0, 0 \rangle = 0 \quad & \text{for} & \quad S=1/2 \; , \\
\label{eq:JLimit2}
- \left\langle \frac{1}{2}, -\frac{1}{2} \left| \op{S}^z \right| \frac{1}{2}, -\frac{1}{2} \right\rangle = \frac{2}{3} \quad & \text{for} & \quad S=1 \; , \\
\label{eq:JLimit3}
- \langle 1, -1 | \op{S}^z | 1, -1 \rangle = \frac{5}{4} \quad & \text{for} & \quad S=3/2 \; .
\end{eqnarray}

\noindent In contrast, in the limit $\rho J \rightarrow \infty$ the impurity contribution to the magnetization $M_{\text{imp}}/g_S \mu_B$ reduces to $- \langle \op{S}^z + \op{s}^z \rangle$ with respect to the above eigenstates
of Hamiltonian (\ref{eq:HEffectiveIso}). This expectation value gives $S-1/2$. The case $S=1/2$ is therefore special since both $M_{\text{imp}}$ and $\mathcal{M}$ go to zero for $\rho J \rightarrow \infty$. Figs.
\ref{fig:MSat_vs_J-isotrop-S1} and \ref{fig:MSat_vs_J-isotrop-S3-2} show that NRG, as a method that is non-perturbative in $J$, can in fact reproduce the limiting values for large coupling strength. \footnote{However, while
being non-perturbative in $J$, NRG still relies on a discretization of the electronic degrees of freedom. For $\rho J > 0.4$ we observe artifacts in our NRG results: $M_{\text{imp}}$ becomes negative for $S=1/2$, whereas
$M_{\text{imp}}$-curves for different magnetic fields eventually cross for $S \geq 1$. In contrast, the impurity magnetization $\mathcal{M}$ does not display any obvious anomalies. Figs. \ref{fig:MSat_vs_J-isotrop-S1} and
\ref{fig:MSat_vs_J-isotrop-S3-2} are basically unaffected by the aforementioned problems.}

Let us now consider the magnetization for intermediate values of $\rho J$. In the special case $S=1/2$ both $M_{\text{imp}}$ and $\mathcal{M}$ are monotonically decreasing functions of the coupling strength that show
similar behavior. We furthermore find that a larger value of the magnetic field $B$ also leads to a larger value of the magnetization. In fact, \emph{all} magnetization curves that we have calculated increase monotonically
with $B$. On the other hand, $M_{\text{imp}}$ and $\mathcal{M}$ display qualitatively different behavior for impurity spin $S \geq 1$. While $M_{\text{imp}}$ is again a monotonically decreasing function of the coupling
strength, $\mathcal{M}$ develops a minimum for all considered magnetic fields.

It is instructive to compare the NRG results with the Bethe Ansatz solution for $M_{\text{imp}}$. According to the standard estimate for the Kondo temperature from Eq. (\ref{eq:TKEstimate}), $k_B T_K/W$ is a monotonically
increasing function of the coupling strength $\rho J$. Furthermore, a larger magnetic field always results in a larger magnetization for the system under consideration. For this reason, an increase of the coupling $\rho J$
leads to a larger Kondo temperature $T_K$ and thus, after rescaling, to a lower effective field and hence a smaller value of $M_{\text{imp}}$. The impurity contribution to the magnetization should  therefore be a monotonically
decreasing function of the coupling strength in the scaling regime. This observation is consistent with our NRG results for $M_{\text{imp}}$ as reported in Figs. \ref{fig:MSat_vs_J-isotrop-S1} and \ref{fig:MSat_vs_J-isotrop-S3-2}.
The vertical lines in the plots mark a coupling strength $\rho J'$ for which we have checked by comparing with the BA solution that $M_{\text{imp}}$ still displays universal behavior. It can thus be seen as a lower bound for
the couplings that can be assigned to the scaling regime. In contrast, the impurity magnetization $\mathcal{M}$ has a minimum for coupling strengths that are smaller than $\rho J'$. We conclude that this behavior of
$\mathcal{M}$ is not compatible with the standard scaling picture as described above.

\section{\label{sec:easyAxis}Impurities with easy axis anisotropy}

We now deal with the case of an impurity with additional easy axis anisotropy (i.e., with anisotropy parameter $D<0$ in Eq. (\ref{eq:HImp})). In this section, emphasis is placed on the field dependence of the
impurity magnetization $\mathcal{M}$, again for the case of equal g-factors. Before considering the full impurity model given by Eq. (\ref{eq:HTotal}), let us briefly recapitulate the magnetic properties of a free
anisotropic spin with Hamiltonian (\ref{eq:HImp}).

For negative anisotropy parameter $D$ and vanishing magnetic field, the groundstate of a spin $S \geq 1$ is a doublet composed of the states with magnetic quantum number $M = \pm S$. In the special
case $S=1/2$, the anisotropy term $D(\op{S}^z)^2$ evaluates to a constant and is thus insignificant for the thermodynamics. The first excited state is a singlet with $M=0$ for $S=1$ and a doublet with
$M=\pm(S-1)$ for all larger spins. It follows that the energy gap between groundstate and first excited state is given by $|D|(2S-1)$. For thermal energies that are small compared to this gap, the zero-field
magnetic susceptibility approximately obeys a Curie law with Curie constant $\langle (\op{S}^z)^2 \rangle = S^2$ (instead of $S(S+1)/3$ for an isotropic spin).

What do we expect for the full impurity model if there is an additional easy axis anisotropy? Since the groundstate doublet of the free anisotropic spin has $\Delta M = 2S > 1$, the two states it is comprised
of are not connected by a single spin-flip, which changes $M$ by 1. Furthermore, for increasing values of $|D|$ the gap in the energy spectrum of the free anisotropic spin progressively suppresses scattering
processes connecting groundstate and first excited state. With the scattering picture in mind, one would thus assume that the Kondo effect is weakened by a negative value of $D$. This is in line with the
simplified picture for the limit $|D|~\rightarrow~\infty$: The anisotropy term $D(\op{S}^z)^2$ then effectively acts as a projection operator onto the groundstate doublet of the impurity with $M=\pm S$ and hence
asymptotically reduces the full Kondo interaction of Eq. (\ref{eq:KondoCoupling}) to an Ising-type coupling. \cite{Zitko2008b} With respect to the impurity magnetization $\mathcal{M}$, there appears to be an
even simpler argument: A larger negative value of the anisotropy parameter $D$ energetically lifts all excited states of the impurity, which have reduced magnetic moment in comparison to the groundstate
doublet. At large $|D|$ one would thus expect that the excited states have less weight in the many-body groundstate of the full impurity model leading to an increased value of $\mathcal{M}$ at zero temperature
for positive magnetic field.


\subsection{\label{subsec:easyAxis_M}Field dependence of the impurity magnetization}

\begin{figure}
\centering{\includegraphics[width=1.0 \linewidth]{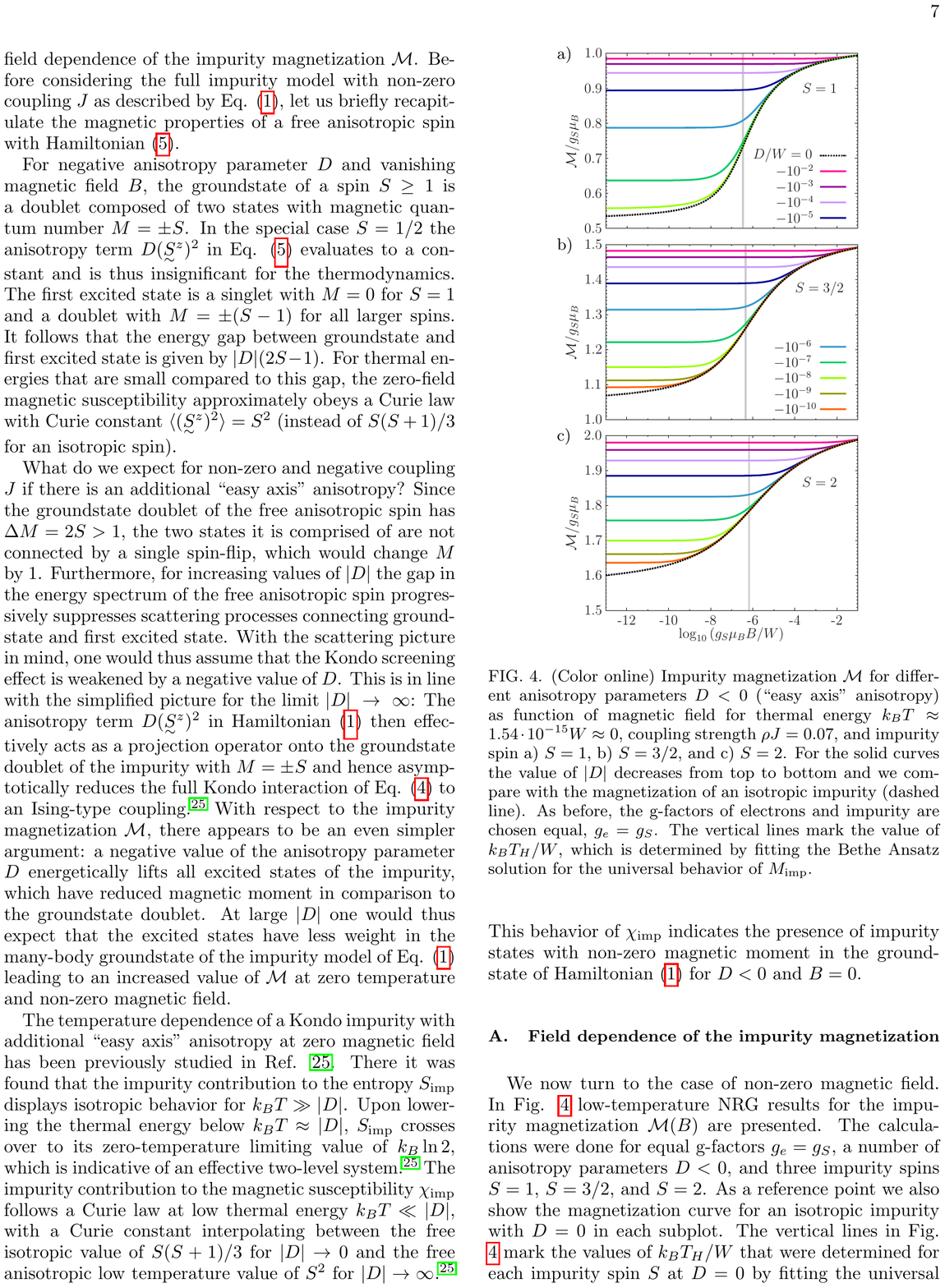}}
%
%
%
\caption{\label{fig:M_vs_B-easyAxis}(Color online) Impurity magnetization $\mathcal{M}$ for different anisotropy parameters $D<0$ (easy axis anisotropy) as function of magnetic field for
$k_B T /W \approx 1.54 \cdot 10^{-15} \approx 0$, coupling strength $\rho J = 0.07$, and impurity spin a) $S=1$, b) $S=3/2$, and c) $S=2$. For the solid curves the value of $|D|$ increases
from bottom to top and we compare with the magnetization of an isotropic impurity (dashed line). As before, equal g-factors of electrons and impurity are assumed. Vertical lines mark the value
of $k_B T_H/W$, which is determined by fitting the universal Bethe Ansatz solution for $M_{\text{imp}}$ with $D=0$. For $\rho J = 0.07$ and $S=2$, we find $k_B T_H / W \approx 6.8 \cdot 10^{-7}$.}
\end{figure}

In Fig. \ref{fig:M_vs_B-easyAxis} low-temperature NRG results for the impurity magnetization $\mathcal{M}(B)$ for impurity spin $S=1,3/2,2$ are presented. We start the discussion of the results
at high magnetic fields and move from there to lower fields. If the Zeeman energy is much larger than the anisotropy parameter, i.e., if $g_S \mu_B B \gg |D|$, nearly isotropic behavior of $\mathcal{M}$
is observed. At smaller fields $g_S \mu_B B \approx |D|$, the impurity magnetization for $D<0$ begins to deviate from the isotropic curve and, for $g_S \mu_B B \ll |D|$, converges to a $D$-dependent
value larger than $g_S \mu_B (S-1/2)$. In the limit of low fields, the impurity magnetization curves for $D<0$ shown in Fig. \ref{fig:M_vs_B-easyAxis} are well described by a linear law:

\begin{equation}
	\mathcal{M}(B) \approx \mathcal{M}^{(0)}(D) \; + \; \gamma(D) \cdot g_S \mu_B B / W \; .
	\label{eq:easyAxis_BZeroLimit}
\end{equation}

\noindent $\mathcal{M}^{(0)}(D)$ thus corresponds to the impurity magnetization in the groundstate of Hamiltonian (\ref{eq:HTotal}) for infinitesimal magnetic field. The low-field behavior for $D<0$ as
described by Eq. (\ref{eq:easyAxis_BZeroLimit}) is different from  that displayed by an isotropic impurity: For $D=0$ and $S \geq 1$, the impurity contribution to the magnetization $M_{\text{imp}}$, which
is proportional to $\mathcal{M}$ in the isotropic case as demonstrated in Sec. \ref{subsec:M_vs_B-isotrop}, approaches the limit of zero magnetic field with slowly decaying logarithmic corrections.
\cite{Tsvelick1983,Andrei1983,HewsonBook}

From the results presented in Fig. \ref{fig:M_vs_B-easyAxis} we conclude that for non-zero magnetic field and $D<0$ a larger value of $|D|$ leads to a larger impurity magnetization $\mathcal{M}$, with an upper
bound for $|D| \rightarrow \infty$ given by the free saturation value $g_S \mu_B S$. This observation is in agreement with the expectations formulated at the beginning of this section. One might therefore say that
an easy axis anisotropy stabilizes the impurity spin.


\begin{figure}
\centering{\includegraphics[width=1.0 \linewidth]{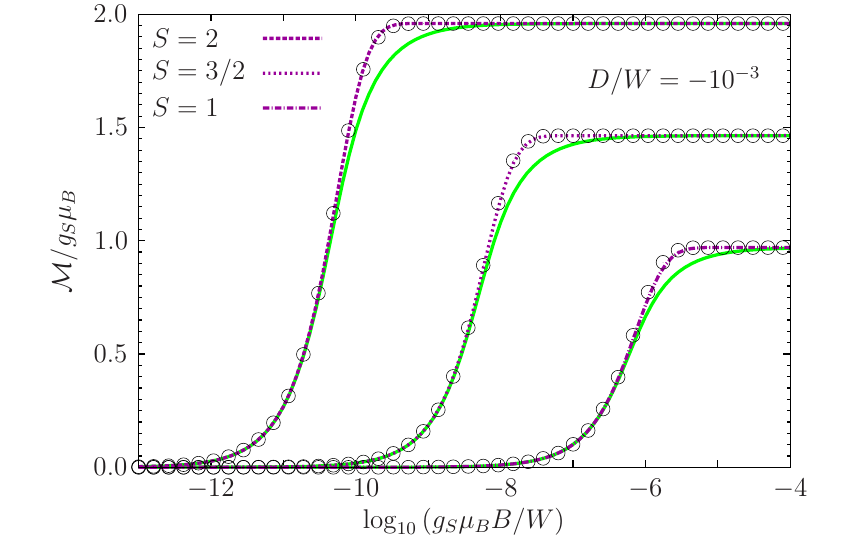}}
%
%
%
\caption{\label{fig:easyAxis_Brillouin-Fit}(Color online) Impurity magnetization $\mathcal{M}$ as function of magnetic field for fixed anisotropy $D/W=-10^{-3}$ and temperature $k_B T/W \approx 1.03 \cdot 10^{-6}$
for $S=1$, $k_B T/W \approx 1.28 \cdot 10^{-8}$ for $S=3/2$, and $k_B T/W \approx 1.58 \cdot 10^{-10}$ for $S=2$. The impurity spin $S$ increases from bottom to top. Note that $\mathcal{M}$ is \emph{not}
saturated for any field in the plot range (cf. Fig. \ref{fig:M_vs_B-easyAxis}). Open symbols represent fits using a rescaled and shifted Brillouin function, $f(x) = \gamma B_S(\eta x)$, and solid (green) lines fits using
a rescaled and shifted Langevin function, $g(x)=\gamma L(\eta x)$.}
\end{figure}

Taking another look at Fig. \ref{fig:M_vs_B-easyAxis} and focusing on the regime of small magnetic fields with Zeeman energy $g_S \mu_B B \ll |D|$, one might be misled to think that there is a saturation of the
impurity magnetization $\mathcal{M}$ (this impression would not occur for an isotropic impurity). This raises the question whether it is possible to approximately describe the field dependence of $\mathcal{M}$
for $g_S \mu_B B \ll |D|$ using a model for a free spin. In the simplest case, such a description could be provided by a Brillouin function $B_S(x)$, which gives the temperature and field dependence of the
magnetization of a free and isotropic spin $S$. As demonstrated in Fig. \ref{fig:easyAxis_Brillouin-Fit} for one value of $D$, it is in fact possible to adequately fit the NRG results for $\mathcal{M}$ from Fig.
\ref{fig:M_vs_B-easyAxis} using a rescaled and shifted Brillouin function, $f(x) = \gamma B_S(\eta x)$ with free parameters $\gamma$ and $\eta$, as long as $k_BT \ll |D|$ and $g_S \mu_B B \ll |D|$. However,
for larger fields the behavior of the impurity magnetization clearly differs from the prediction of a Brillouin function, as seen in Fig. \ref{fig:M_vs_B-easyAxis}. The ratio of $|D|$ and $k_BT_H$ (or alternatively
$k_BT_K$) determines the ``apparent saturation value'' of $\mathcal{M}$ and thus the parameter $\gamma$. In contrast to a fit with a modified Brillouin function, a classical description using a rescaled and
shifted Langevin function, $g(x) = \gamma L(\eta x)$ with $L(x) = \lim_{S \rightarrow \infty}{B_S(x)}$, does not work well for magnetic fields close to the ``saturation field'' (cf. the solid lines in Fig.
\ref{fig:easyAxis_Brillouin-Fit}), as is to be expected for a quantum mechanical system with low spin. Nevertheless, a fit using $g(x)$ can produce reasonable results for fields that are small compared to the
``saturation field''.

The results depicted in Fig. \ref{fig:easyAxis_Brillouin-Fit} might be of importance for an experimental study of a system that is (approximately) described by Hamiltonian (\ref{eq:HTotal}) with a strong easy
axis anisotropy. It is then conceivable that a measurement of the magnetization for magnetic fields that can be realistically produced in an experiment (depending on the value of $D$, fields with
$g_S \mu_B B \approx |D|$ might not be obtainable) does not allow to distinguish between the behavior of an anisotropic impurity spin and that of a free spin. Such a scenario seems more likely if the
experimental control over the g-factor and the absolute magnitude of the magnetization is limited, and if $|D|$ is large compared to $k_BT_K$ so that the ``apparent saturation value'' of the impurity
magnetization lies close to the free saturation value $g_S \mu_B S$.

\subsection{\label{subsec:easyAxis_Mimp}Impurity contribution to the magnetization and the susceptibility}

As for an isotropic impurity (cf. Sec. \ref{subsec:M_vs_B-isotrop}), we have analyzed the connection between the impurity magnetization $\mathcal{M}$, which is shown in Fig. \ref{fig:M_vs_B-easyAxis}, and the impurity
contribution to the magnetization $M_{\text{imp}}$ (not shown) for $D<0$. It is found that the relation between both quantities is the same as in the isotropic case, i.e., $\mathcal{M} = \alpha M_{\text{imp}}$ with a
proportionality factor $\alpha$ that is independent of the anisotropy parameter $D$ when taking into account the precision of the results from Table \ref{tab:T_H-isotropic}. For impurity spin $S=2$ and coupling strength
$\rho J=0.07$ the obtained value of the proportionality factor is $\alpha = 1.03701(3)$.

We have furthermore investigated the relationship between $M_{\text{imp}}$ at non-zero magnetic field and the impurity contribution to the susceptibility $\chi_{\text{imp}}$ at zero field. At low temperature
$k_BT \ll |D|$, $\chi_{\text{imp}}$ obeys a Curie law with a Curie constant interpolating between the free isotropic value of $S(S+1)/3$ for $|D| \rightarrow 0$ and the free anisotropic low temperature value
of $S^2$ for $|D| \rightarrow \infty$. \cite{Zitko2008b} It turns out that there is a simple relation between the Curie constant and the zero-temperature magnetization $M_{\text{imp}}$ for small magnetic fields
$g_S \mu_B B \ll |D|$:

\begin{equation}
	\left. \frac{k_B T \chi_{\text{imp}}}{(g_S \mu_B)^2} \right|_{B=0, \; k_BT \ll |D|} \approxeq \left. \left( \frac{M_{\text{imp}}}{g_S \mu_B} \right)^2 \right|_{\tilde{h} \ll |D|, \; k_BT \ll \tilde{h}}
	\label{eq:chiMimpRelation}
\end{equation}

\noindent with $\tilde{h} = g_S\mu_BB$. The relative deviation between left hand and right hand side of Eq. (\ref{eq:chiMimpRelation}), as determined by NRG calculations for all parameter combinations used in Fig.
\ref{fig:M_vs_B-easyAxis}, is less than 1 \textperthousand. The relationship between zero-field susceptibility and magnetization expressed by Eq. (\ref{eq:chiMimpRelation}) is actually the same as for a doublet composed of
states with magnetic quantum numbers $\pm M$. In particular, a free spin with easy axis anisotropy effectively reduces to such a doublet at low temperature $k_BT \ll |D|$, as discussed at the beginning of this
section.

In summary, the following picture of the low-temperature properties of a Kondo impurity with easy axis anisotropy is obtained: The impurity effectively acts as a two-level system with residual entropy
$S_{\text{imp}} = k_B \ln{2}$, \cite{Zitko2008b} that splits up in a small magnetic field (leading to $S_{\text{imp}}=0$) and displays an unusual groundstate magnetization indicative of a so-called
``fractional spin'' \cite{Zitko2008b}. Despite this special property, the combination of $\chi_{\text{imp}}$ and $M_{\text{imp}}$ shows that the magnetic response at low temperature and field still
resembles that of an ordinary magnetic doublet.

\section{\label{sec:hardAxis} Impurities with hard axis anisotropy}

We now investigate how an additional hard axis anisotropy ($D>0$) affects the magnetic properties of the impurity. To lay the foundations for a study of the full impurity problem, we first discuss the
magnetic field dependence of the magnetization for a free anisotropic spin described by Hamiltonian (\ref{eq:HImp}) with $D>0$.

\begin{figure}
\centering{\includegraphics[width=1.0 \linewidth]{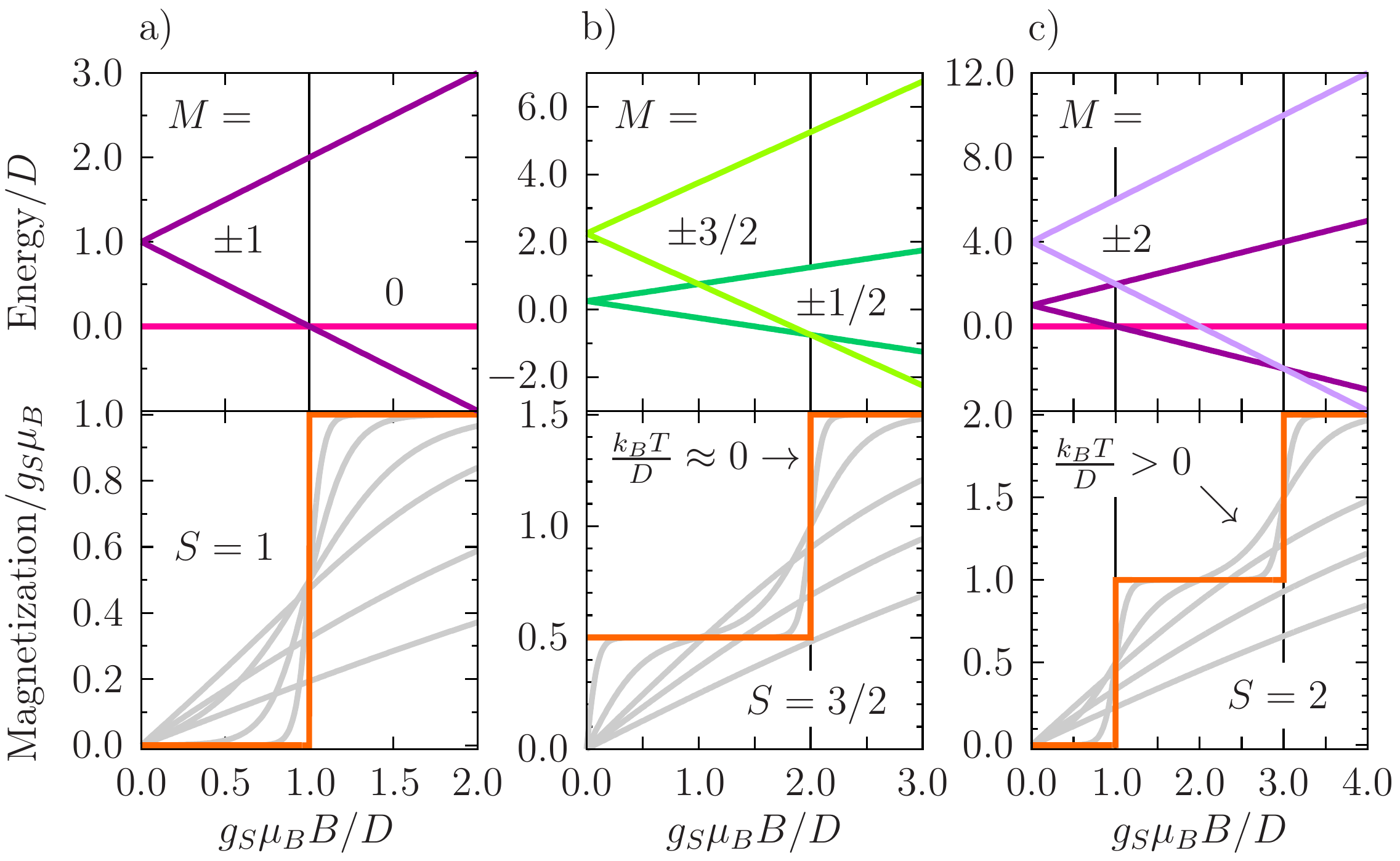}}
%
%
\caption{\label{fig:M_vs_B-hardAxis-freeSpin}(Color online) Energy eigenvalues with magnetic quantum numbers $M$ (upper panels) and magnetization (lower panels) as function of magnetic field for
an anisotropic spin, described by Hamiltonian (\ref{eq:HImp}) with $D>0$ (hard axis anisotropy), with a) $S=1$, b) $S=3/2$, and c) $S=2$. For the magnetization curves and fields larger than the
respective saturation field, temperature increases from top to bottom.}
\end{figure}

For positive $D$ and $B=0$, the eigenvalues of Hamiltonian (\ref{eq:HImp}) are energetically ordered according to the absolute value of their respective magnetic quantum number $M$. Depending
on the spin $S$, the groundstate is thus either a singlet with $M=0$ (for integer $S$) or a doublet with $M=\pm 1/2$ (for half-integer $S$). In either case, the rest of the energy spectrum is comprised
of doublets with magnetic quantum numbers $\pm M$ and $1/2 < M \leq S$. The energy gap $\Delta_{|M|}$ between a level with quantum number $M$ and the next higher-lying doublet is given by
$\Delta_{|M|} = (2|M|+1) D$. As a consequence of the magnetic field dependence of the eigenvalues described by the Zeeman term in Eq. (\ref{eq:HImp}), $n$ groundstate level crossings occur for
positive magnetic fields, with $n=S$ for integer spin and $n=S-1/2$ for half-integer spin. At the field $B_M = \Delta_{|M|}/g_S \mu_B$, the magnetic quantum number of the groundstate abruptly
changes from $-M$ to $-(M+1)$ and hence the zero-temperature magnetization curve displays a discontinuous step. This effect is illustrated in Fig. \ref{fig:M_vs_B-hardAxis-freeSpin} for spin
$S=1,3/2,$ and 2. Finite temperature smears out the magnetization steps and renders them continuous. As the low-energy situation is the same in the vicinity of each groundstate level crossing,
so is the effect of moderate temperature (cf. Fig. \ref{fig:M_vs_B-hardAxis-freeSpin} c).

\subsection{\label{subsec:hardAxisMagn}Magnetic field dependence of the impurity magnetization}

\begin{figure}
\centering{\includegraphics[width=1.0 \linewidth]{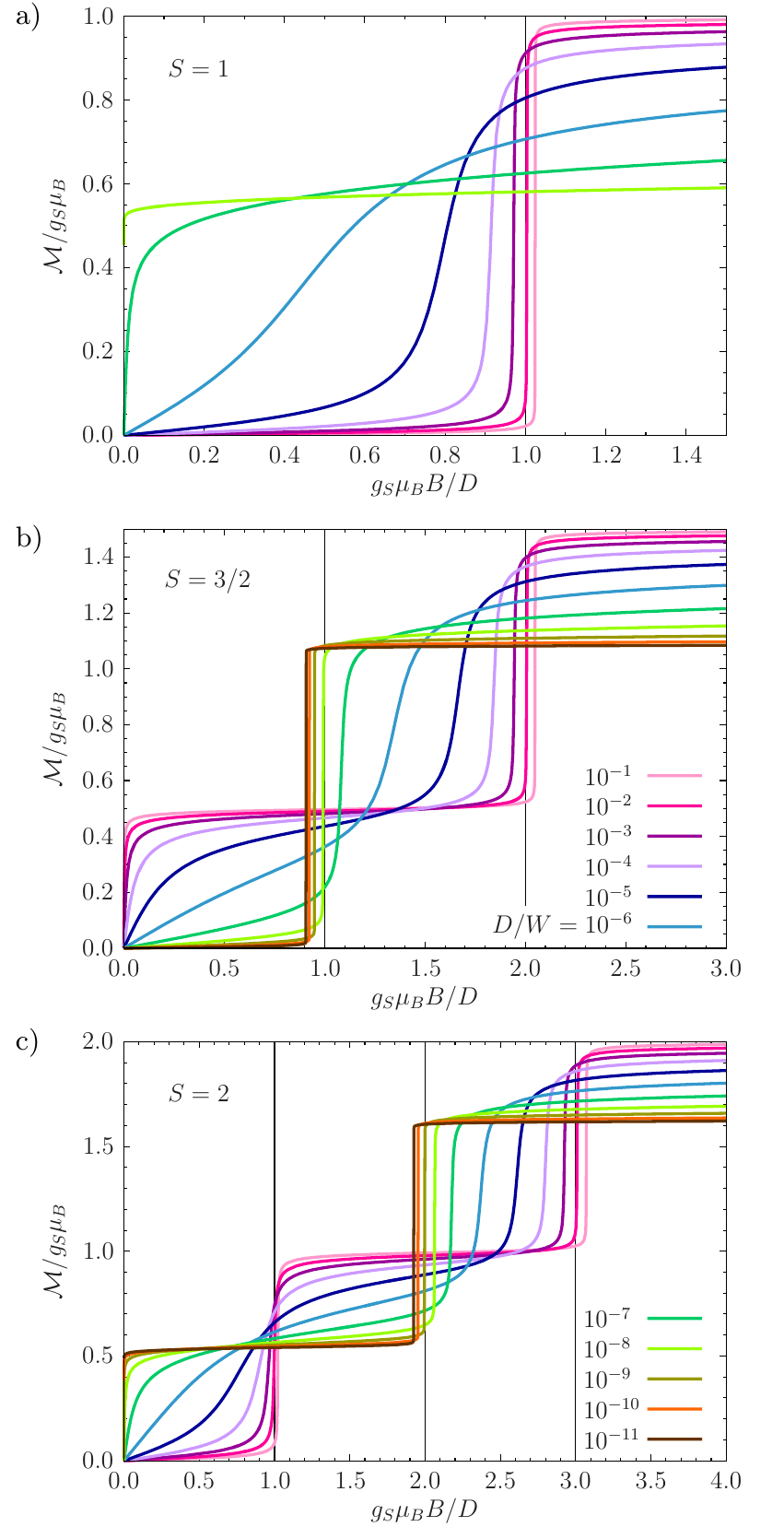}}
%
%
%
%
\caption{\label{fig:M_vs_BD-hardAxis}(Color online) Impurity magnetization $\mathcal{M}$ for different anisotropy parameters $D>0$ as function of magnetic field for temperature
$k_B T/W \approx 1.54 \cdot 10^{-15} \approx 0$, coupling strength $\rho J = 0.07$, equal g-factors, and impurity spin a) $S=1$, b) $S=3/2$, and c) $S=2$. The value of $D$ increases
from bottom to top or from left to right, respectively. Note the use of a linear  magnetic field scale and the rescaling of the Zeeman energy $g_S \mu_B B$ by $D$.}
\end{figure}

We begin with the discussion of the magnetic field dependence of the impurity magnetization $\mathcal{M}(B)$ for equal g-factors and quasi-vanishing temperature $T \approx 0$. Magnetization curves for
impurity spin $S=1,3/2,$ and 2, and several values of the anisotropy parameter $D>0$ are shown in Fig. \ref{fig:M_vs_BD-hardAxis}. Since the coupling strength $\rho J$, and thus the isotropic energy scale
$k_BT_H$ according to Table \ref{tab:T_H-isotropic}, is kept constant, the ratio $k_BT_H/D$ is varied. Note that a linear magnetic field scale is used in Fig. \ref{fig:M_vs_BD-hardAxis} to allow for an easy
comparison with the results for a free anisotropic spin from Fig. \ref{fig:M_vs_B-hardAxis-freeSpin}. 

For an interpretation of the results for $\mathcal{M}(B,T \approx 0)$, let us first consider the two limiting cases in which $D$ is either small or large compared to $k_BT_H$. In the following, imagine that we
move from large magnetic fields to lower fields. If $D$ is large, then little isotropic Kondo screening can occur before the anisotropy becomes effective. As a \emph{guideline}, we might thus think of the energy
spectrum of a free anisotropic spin $S$. On the other hand, for small $D$ significant isotropic Kondo screening can take place before the magnetic field reaches the energy scale defined by the anisotropy, so
that it eventually becomes more appropriate to think of the energy levels of a free anisotropic spin $S-1/2$. As illustrated in Fig. \ref{fig:M_vs_B-hardAxis-freeSpin}, groundstate level crossings occur for a spin
with hard axis anisotropy at certain fields and give rise to steps in the zero-temperature magnetization curve. For small and large anisotropy $D$, the impurity magnetization $\mathcal{M}(B,T \approx 0)$ also
displays sharp, step-like features that are surrounded by magnetic field domains in which $\mathcal{M}$ only slowly increases with $B$ (``pseudo-plateaus''). However, due to the energy continuum of electronic
states, these sharp features stay continuous even in the limit of zero temperature. The number of steps and their position relative to the field $D/g_S \mu_B$ depends on the impurity spin $S$ for large $D$ and
on the residual spin $S-1/2$ for small $D$, respectively. The case $S=1$ is special because the residual spin-1/2 cannot have uniaxial anisotropy. The single step which exists for $S=1$ at large $D$ therefore
disappears for smaller values of $D$. With respect to the energy scale imposed by the anisotropy, the steps in the impurity magnetization become well-defined for small and large $D$. In addition, the pseudo-plateaus
become flatter. Fig. \ref{fig:M_vs_BD-hardAxis}~c furthermore suggests that the two steps appearing in $\mathcal{M}(B,T \approx 0)$ for $S=2$ and large $D$ have different width. We are going to discuss the
aspect of step width in more detail in Sec. \ref{subsec:fieldInducedKondoEffect}. In particular, it will be shown that an impurity magnetization step is steeper if it occurs at larger field. It turns out that a standard
$z$-averaging of the NRG results introduces artifacts into the magnetization steps for large anisotropy $D$. This problem is investigated in more detail in the context of an effective model at the end of Sec.
\ref{subsec:fieldInducedKondoEffect}. The plots shown in Fig. \ref{fig:M_vs_BD-hardAxis} are not visibly affected by this numerical shortcoming.

The position of the steps in the impurity magnetization curves for both small and large anisotropy $D$ seems interesting. Figs. \ref{fig:M_vs_BD-hardAxis} b and c show that for small $D$ and impurity spin $S=3/2$
and $S=2$ a step occurs at a magnetic field which is smaller than the corresponding level crossing field for a free anisotropic spin $S-1/2$. In contrast, for large $D$ and all three impurity spins considered in Fig.
\ref{fig:M_vs_BD-hardAxis}, each impurity magnetization step is found at a field exceeding the corresponding level crossing field for a spin $S$. One might wonder whether the half-bandwidth $W$ of the electrons
has an impact on these two effects. To investigate this question for the case of small anisotropy, we have calculated additional magnetization curves for impurity spin $S=3/2$ and $S=2$ with decreasing coupling
strength $\rho J$ (0.09, 0.07, and 0.05). Since $\rho = 1/2W$, a reduction of $\rho J$ can be interpreted as an increase of the half-bandwidth $W$ for fixed $J$. For each value of the coupling strength the anisotropy
parameter was chosen according to Table \ref{tab:T_H-isotropic} so as to give a constant ratio $k_BT_H/D = 1000$. With this choice of $D$, a reduction of $\rho J$ leads to a shift of the step in the impurity magnetization
curve towards smaller fields \emph{relative to} $D/g_S \mu_B$. This suggests that the effect observed for small $D$ is not bandwidth-related. The question at which fields impurity magnetization steps occur for large
anisotropy is considered in chapter \ref{subsec:fieldInducedKondoEffect}.

The limiting cases of small and large anisotropy $D$ are connected by a regime with partial isotropic Kondo screening in which it is not possible to exclusively think in terms of the impurity spin $S$ or an residual
spin $S-1/2$. Impurity magnetization steps are broadened in this regime with respect to the energy scale $D$ and, upon reducing the anisotropy parameter, move towards lower fields relative to $D/g_S \mu_B$
(see Fig. \ref{fig:M_vs_BD-hardAxis} b). In addition, one step disappears for $D \approx k_BT_H$ in case of integer impurity spin $S$ (cf. Fig. \ref{fig:M_vs_BD-hardAxis} c). Isotropic Kondo screening effectively
changes the impurity spin $S$ from integer to half-integer and vice versa. \cite{Zitko2008b} Depending on the ratio $k_BT_H/D$ and the resulting degree of isotropic Kondo screening, different behavior of the
impurity magnetization is therefore observed, in particular for fields $g_S \mu_B B \lesssim D$: While there is little magnetic response for an effective integer spin (at zero temperature, there is none at all for a free
anisotropic integer spin as shown in Fig. \ref{fig:M_vs_B-hardAxis-freeSpin}), we have larger impurity magnetization $\mathcal{M} \approx g_S \mu_B/2$ for effective half-integer spin. However, in the latter case an
additional pseudo-spin-1/2 Kondo effect occurs \cite{Zitko2008b} which leads to a suppression of the impurity magnetization for magnetic fields $g_S \mu_B B \ll D$.


\begin{figure}
\centering{\includegraphics[width=1.0 \linewidth]{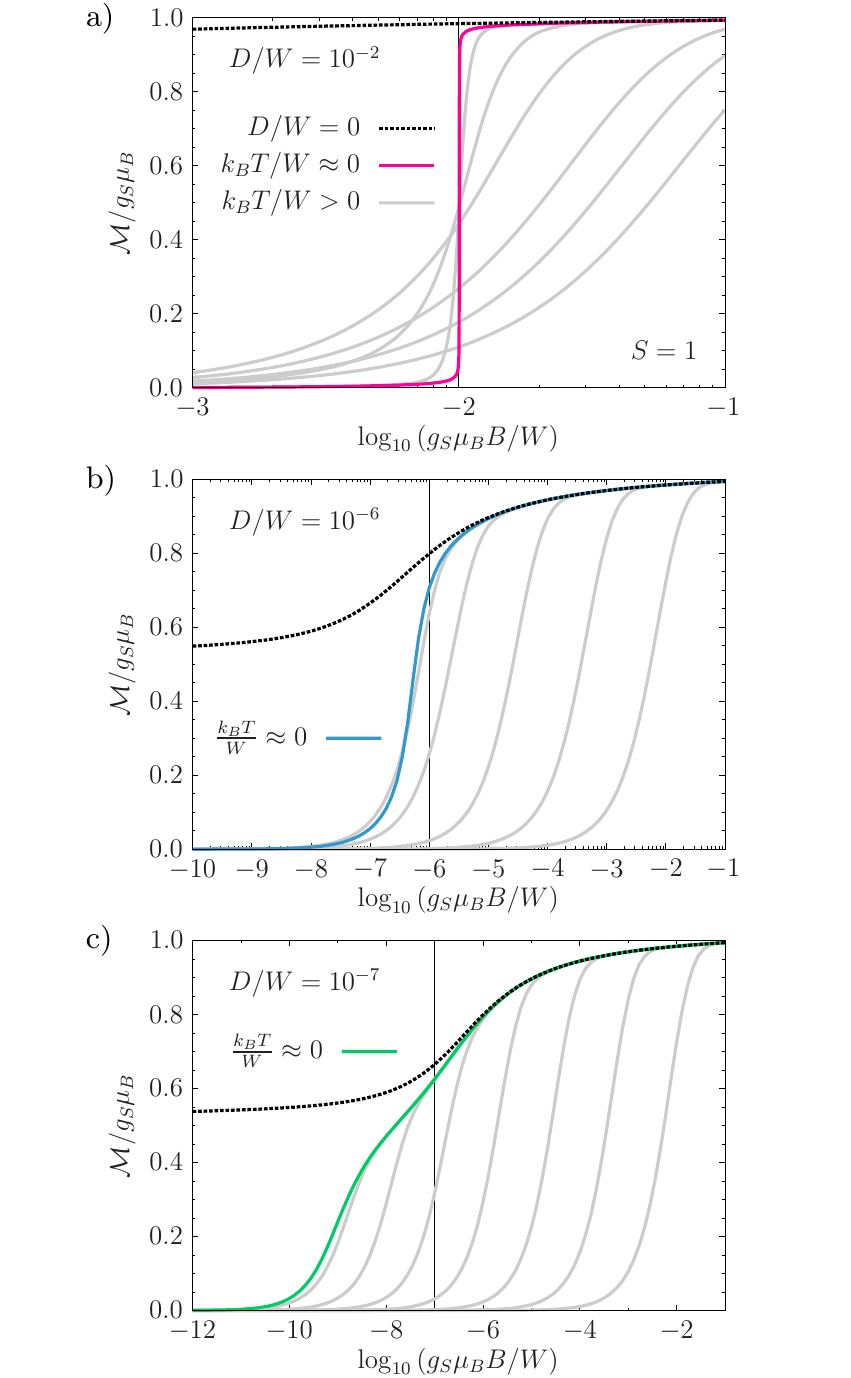}}
%
%
%
%
\caption{\label{fig:M_vs_B-hardAxis-S1}(Color online) Impurity magnetization $\mathcal{M}$ as function of magnetic field for impurity spin $S=1$ and anisotropy parameter a) $D/W=10^{-2}$, b) $D/W=10^{-6}$,
and c) $D/W=10^{-7}$ (solid lines, cf. Fig. \ref{fig:M_vs_BD-hardAxis}). Dashed lines show $\mathcal{M}(B)$ for the isotropic case and $k_B T/W \approx 1.54 \cdot 10^{-15} \approx 0$. For the light gray lines,
the approximate value of $k_BT/W$ increases from left to right from a) $4.35 \cdot 10^{-4}$ to $6.10 \cdot 10^{-2}$, b) $1.99 \cdot 10^{-7}$ to $6.79 \cdot 10^{-3}$, and c) $4.73 \cdot 10^{-10}$ to
$6.79 \cdot 10^{-3}$. Thin vertical lines indicate the respective Zeeman energy $g_S \mu_B B = D$. For the chosen coupling strength $\rho J=0.07$ and $D=0$, we have $k_BT_H/W \approx 3.39 \cdot 10^{-7}$
according to Table \ref{tab:T_H-isotropic}.}
\end{figure}

\begin{figure}
\centering{\includegraphics[width=1.0 \linewidth]{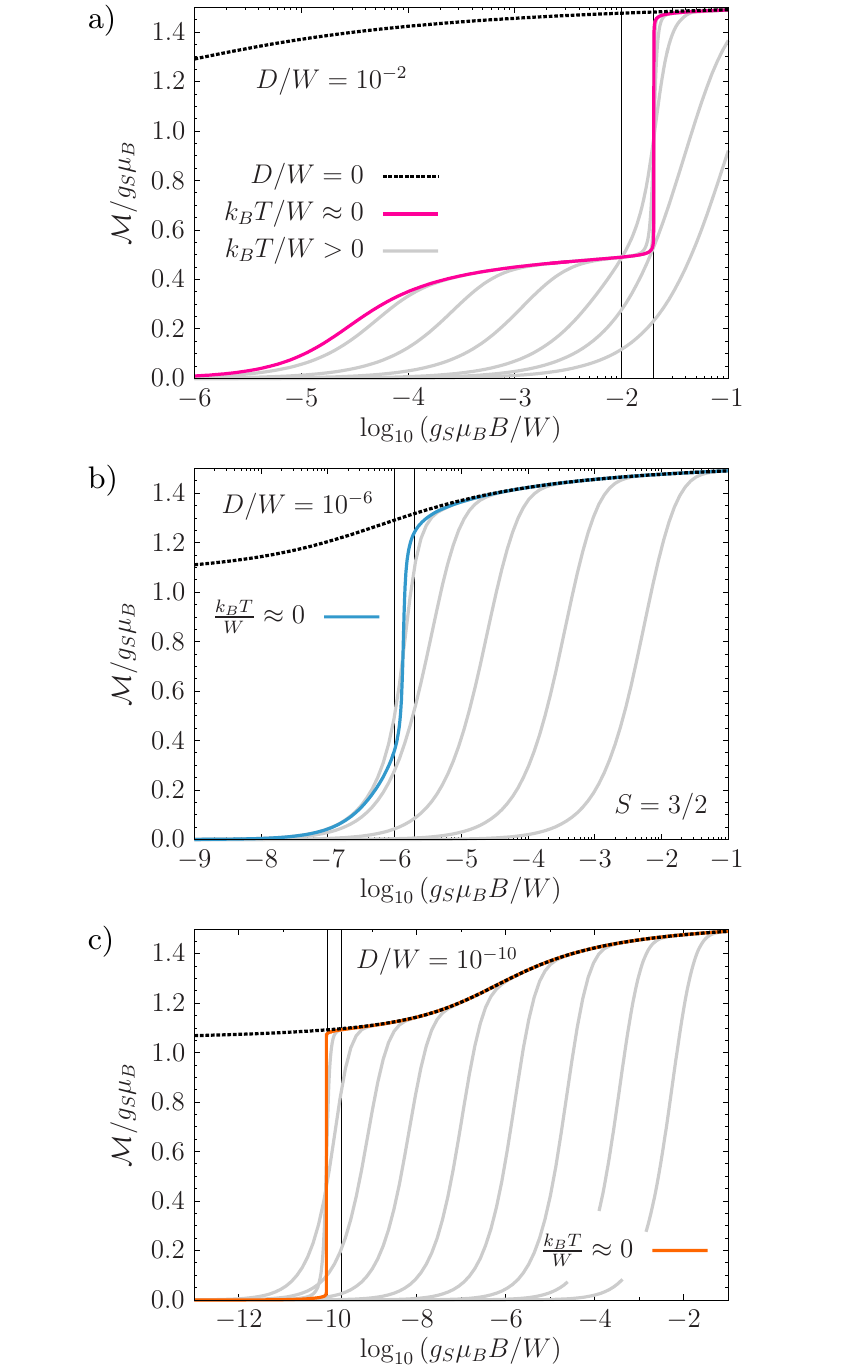}}
%
%
%
%
\caption{\label{fig:M_vs_B-hardAxis-S3-2}(Color online) Impurity magnetization $\mathcal{M}$ as function of magnetic field for impurity spin $S=3/2$ and anisotropy parameter a) $D/W=10^{-2}$, b)
$D/W=10^{-6}$, and c) $D/W=10^{-10}$ (solid lines, cf. Fig. \ref{fig:M_vs_BD-hardAxis}). As before, dashed lines show $\mathcal{M}(B)$ for the isotropic case and $k_B T/W \approx 0$. For the light
gray lines, the approximate value of $k_BT/W$ increases from left to right from a) $1.61 \cdot 10^{-5}$ to $1.05 \cdot 10^{-1}$, b) $3.45 \cdot 10^{-7}$ to $6.79 \cdot 10^{-3}$, and c) $1.01 \cdot 10^{-11}$
to $6.79 \cdot 10^{-3}$. Magnetic fields satisfying $g_S \mu_B B = n D$ with $n=1,2$ are highlighted by thin vertical lines. According to Table \ref{tab:T_H-isotropic}, $k_BT_H/W \approx 4.55 \cdot 10^{-7}$
for $\rho J=0.07$ and $D=0$.}
\end{figure}

\begin{figure}
\centering{\includegraphics[width=1.0 \linewidth]{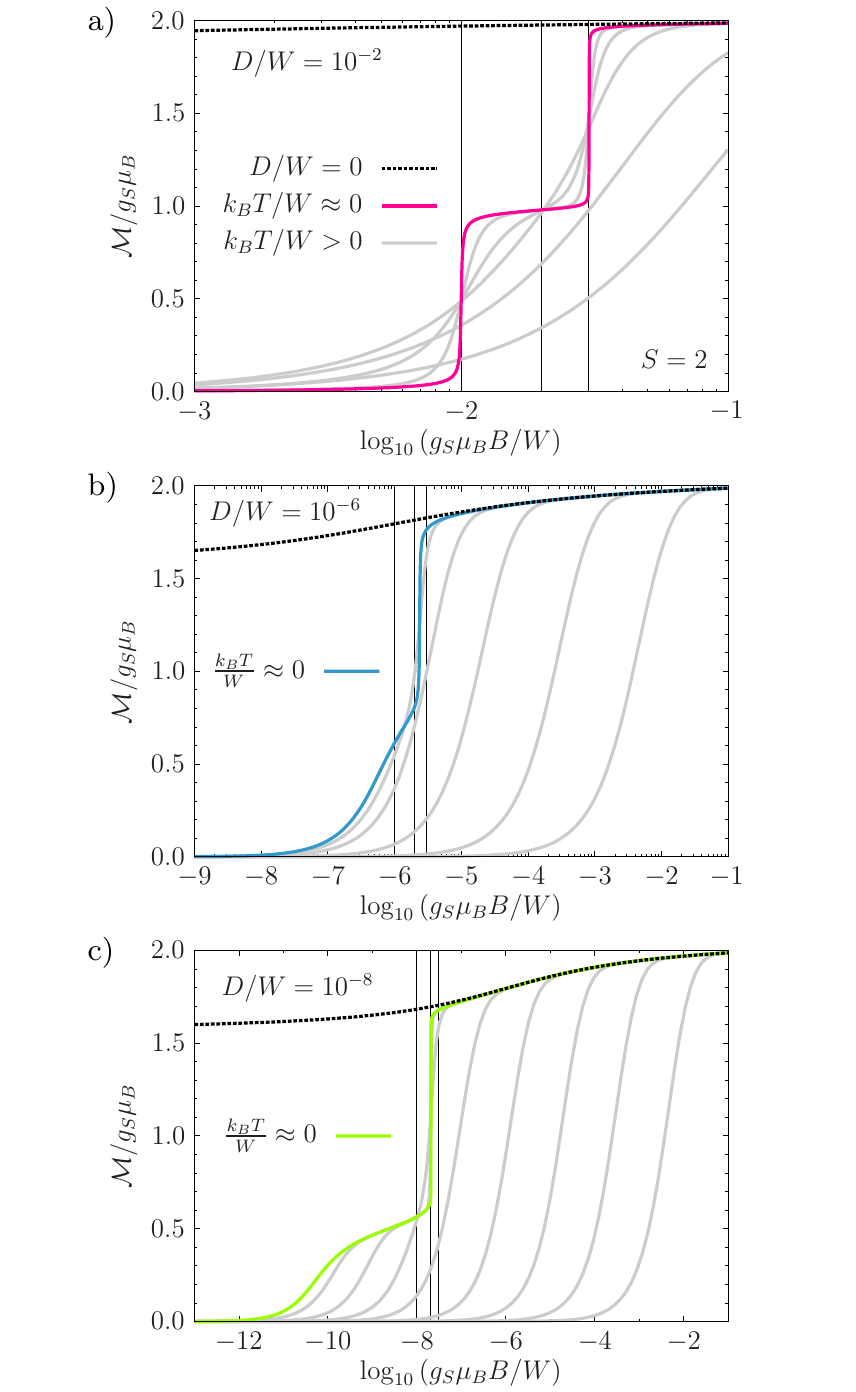}}
%
%
%
%
\caption{\label{fig:M_vs_B-hardAxis-S2}(Color online) Impurity magnetization $\mathcal{M}$ as function of magnetic field for impurity spin $S=2$ and anisotropy parameter a) $D/W=10^{-2}$, b) $D/W=10^{-6}$,
and c) $D/W=10^{-8}$ (solid lines, cf. Fig. \ref{fig:M_vs_BD-hardAxis}). Dashed lines again show $\mathcal{M}(B)$ for the isotropic case and $k_B T/W \approx 0$. For the light gray lines, the approximate value of
$k_BT/W$ increases from left to right from a) $7.54 \cdot 10^{-4}$ to $1.05 \cdot 10^{-1}$, b) $3.45 \cdot 10^{-7}$ to $6.79 \cdot 10^{-3}$, and c) $5.26 \cdot 10^{-11}$ to $6.79 \cdot 10^{-3}$. Thin vertical lines
indicate magnetic fields satisfying $g_S \mu_B B = n D$, with $n=1,2,3$. For $\rho J=0.07$ and $D=0$, we obtain $k_BT_H/W \approx 6.8 \cdot 10^{-7}$.}
\end{figure}

In order to study the three different anisotropy regimes ($k_BT_H/D \ll 1$, $k_BT_H/D \approx 1$, and $k_BT_H/D \gg 1$) in more detail, selected impurity magnetization curves from Fig. \ref{fig:M_vs_BD-hardAxis}
are replotted using a logarithmic magnetic field scale (see Figs. \ref{fig:M_vs_B-hardAxis-S1}, \ref{fig:M_vs_B-hardAxis-S3-2}, and \ref{fig:M_vs_B-hardAxis-S2}). This brings out effective spin-1/2 Kondo effects
more clearly and allows for a better comparison with the behavior of an isotropic impurity. Figs. \ref{fig:M_vs_B-hardAxis-S1}, \ref{fig:M_vs_B-hardAxis-S3-2}, and \ref{fig:M_vs_B-hardAxis-S2} furthermore
demonstrate the effect of finite temperature on the impurity magnetization for the case of hard axis anisotropy.

At large magnetic fields $g_S \mu_B B \gg D$, the magnetization curve $\mathcal{M}(B,T \approx 0)$ for an anisotropic impurity closely resembles the isotropic result. For lower fields $g_S \mu_B B \gtrsim D$, the
anisotropy eventually becomes effective and the impurity magnetization begins to deviate from the isotropic curve. In particular, $\mathcal{M}(B,T \approx 0)$ displays a linear dependence on $B$ for magnetic fields
that are small compared to all relevant energy scales. If temperature is high so that $k_BT \gg D$, then $\mathcal{M}(B)$ is suppressed for magnetic fields of the order of and smaller than $k_BT/g_S \mu_B$. On the
other hand, judging by the relative deviation from the zero-temperature curve, finite temperature has negligible effect on the impurity magnetization if the thermal energy falls into the energy regime in which
$\mathcal{M}(B, T \approx 0) \propto B$ (such a temperature independence is known for an isotropic impurity with $S=1/2$, as discussed in Sec. \ref{subsec:M_vs_B-isotrop}). In this regard, $\mathcal{M}$ differs from
the magnetization of a free anisotropic spin for which non-zero temperature is always relevant. If the impurity magnetization curve features steps, they are smeared out for sufficiently high temperature. In case there is
more than one step in $\mathcal{M}(B,T \approx 0)$ (cf. Fig. \ref{fig:M_vs_B-hardAxis-S2} a), then, as a further difference compared to a free anisotropic spin, non-zero temperature has unequal effect on the different
steps. We come back to the last two observations in Sec. \ref{subsec:fieldInducedKondoEffect}.

Ref. \onlinecite{Zitko2008b} arrives at the conclusion that an impurity with additional hard axis anisotropy undergoes anisotropic spin-1/2 Kondo screening if higher-lying impurity states are frozen out so that the
impurity is effectively reduced to a spin-1/2 doublet. For $g_S \mu_B B \ll D$ and low temperature, the impurity magnetization is therefore either suppressed if the impurity is effectively reduced to a singlet (cf.
Figs. \ref{fig:M_vs_B-hardAxis-S1} a, \ref{fig:M_vs_B-hardAxis-S3-2} c, and \ref{fig:M_vs_B-hardAxis-S2} a), or it reflects the magnetic response of a Kondo screened spin-1/2 doublet (cf. Figs.
\ref{fig:M_vs_B-hardAxis-S1} c, \ref{fig:M_vs_B-hardAxis-S3-2} a, and \ref{fig:M_vs_B-hardAxis-S2} c). For small $D$ and integer impurity spin, the corresponding Kondo temperature $T_K^*$ decreases faster
than linear in $D$ upon reducing the anisotropy (it even drops exponentially fast in the special case $S=1$).\cite{Zitko2008b} As a result, the pseudo-plateau that appears for $S=2$ at magnetic fields
$g_S \mu_B B \lesssim 2D$ (see Figs. \ref{fig:M_vs_B-hardAxis-S2} c and \ref{fig:M_vs_BD-hardAxis} c) becomes flatter and broader when decreasing $D$, whereas a reduction of $D$ eventually leads to
quasi-isotropic behavior over the whole considered magnetic field range for $S=1$ (see Figs. \ref{fig:M_vs_B-hardAxis-S1} c and \ref{fig:M_vs_BD-hardAxis} a). On the other hand, for large and increasing
anisotropy $D$ and half-integer impurity spin, the observed screening effect is increasingly well described by an exchange-anisotropic $S=1/2$-Kondo model, whose Kondo temperature has a value much
smaller than $W/k_B$. \cite{Zitko2008b} This means that the pseudo-plateau occuring for impurity spin $S=3/2$ and fields $g_S \mu_B B \lesssim 2D$ (see Figs. \ref{fig:M_vs_B-hardAxis-S3-2} a and
\ref{fig:M_vs_BD-hardAxis} b) becomes more pronounced for larger anisotropy $D$. As a final remark, Fig. \ref{fig:M_vs_B-hardAxis-S2} a once again shows the different width of the two impurity magnetization
steps for $S=2$ and large $D$.

\subsection{\label{subsec:hardAxisElectronGFactor}Influence of the electron g-factor on $\mathcal{M}$ and the connection between $\mathcal{M}$ and $M_{\text{imp}}$}

\begin{figure}
\centering{\includegraphics[width=1.0 \linewidth]{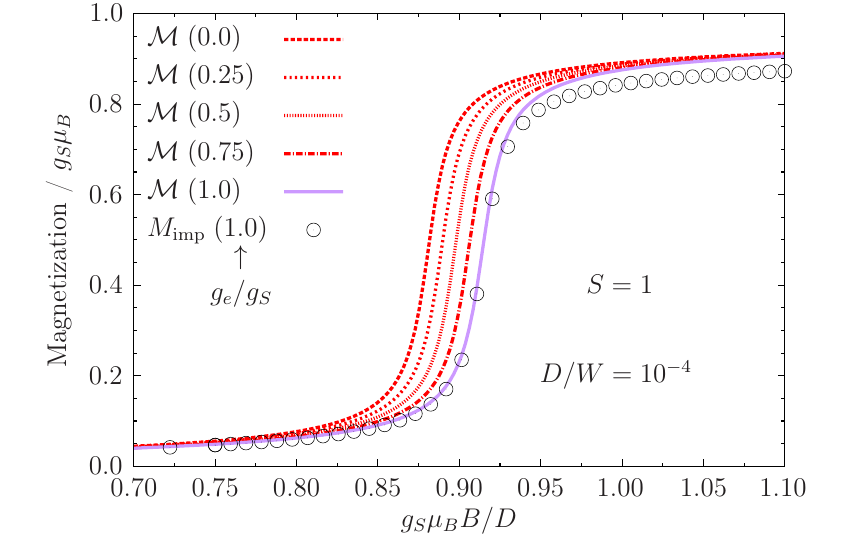}}
%
%
%
%
\caption{\label{fig:M_vs_B-hardAxis-S1-gRatio}(Color online) Impurity magnetization $\mathcal{M}$ as function of magnetic field for impurity spin $S=1$, $D/W=10^{-4}$, coupling strength $\rho J=0.07$, and several
values of the g-factor ratio $g_e/g_S$ (cf. Fig. \ref{fig:M_vs_BD-hardAxis} a for $g_e=g_S$). For equal g-factors, the impurity contribution to the magnetization $M_{\text{imp}}$ is shown, too. The temperature is
$k_BT/W \approx 1.54 \cdot 10^{-15} \approx 0$ in all cases.}
\end{figure}

We now investigate how the impurity magnetization $\mathcal{M}(B, T \approx 0)$ is affected by a non-zero electron g-factor corresponding to a positive ratio $g_e/g_S > 0$. Because of the sharp features that are
found in the magnetization curves, the case of hard axis anisotropy seems well suited to study the influence of non-zero $g_e$ (alternatively one could examine the effect of the electron g-factor on the basis of the
linear magnetic field dependence of $\mathcal{M}(B)$ for small $B$). As an example, Fig. \ref{fig:M_vs_B-hardAxis-S1-gRatio} shows magnetization curves $\mathcal{M}(B,T \approx 0)$ for impurity spin $S=1$,
moderately large $D$, and several g-factor ratios interpolating between a local field ($g_e = 0$) and a bulk field ($g_e=g_S$).

The results presented in Fig. \ref{fig:M_vs_B-hardAxis-S1-gRatio} demonstrate that a positive electron g-factor effectively causes a rescaling of the magnetic field: A ratio $g_e/g_S > 0$ shifts the impurity magnetization
curve for $g_e=0$ towards larger fields and thus reduces $\mathcal{M}(B,T \approx 0)$, which is a monotonically increasing function of $B$, for a fixed magnetic field value. Using the notation $\mathcal{M}(B,T,g_S,g_e)$,
this statement can be expressed as:

\begin{eqnarray}
	\label{eq:MgRelation1} \mathcal{M}(B,T \approx 0, g_S, 0) & = & \mathcal{M}(\widetilde{B}, T \approx 0, g_S, g_e) \; , \\
	\label{eq:MgRelation2} \widetilde{B} & = & \eta\left(\rho J, \frac{g_e}{g_S}\right) B \; ,
\end{eqnarray}

\noindent with a rescaling factor $\eta(\rho J, g_e/g_S) \geq 1$ for $g_e/g_S \geq 0$ that depends on the coupling strength $\rho J$. Taking the magnetization curve for a local field as reference, we may therefore state
that the impurity effectively ``feels'' a smaller magnetic field if there is also a Zeeman term for the electrons in Hamiltonian (\ref{eq:HTotal}).

For a bulk field, Fig. \ref{fig:M_vs_B-hardAxis-S1-gRatio} additionally shows the impurity contribution to the magnetization $M_{\text{imp}}$. Note that $\mathcal{M}$ and $M_{\text{imp}}$ are equal by definition for a
local field. As previously found in the isotropic (cf. Sec. \ref{subsec:M_vs_B-isotrop}) and easy axis case (cf. Sec. \ref{subsec:easyAxis_M}), the numerical results strongly suggest a proportionality
$\mathcal{M}(B, T \approx 0)=\alpha(\rho J) M_{\text{imp}}(B, T \approx 0)$ for fields $g_S \mu_B B \ll W$. A study of the connection between $\mathcal{M}$ and $M_{\text{imp}}$ for anisotropy parameters $0 < D \ll W$
and impurity spin $S=1,3/2,$ and 2 reveals that the average values for $\alpha(0.07)$ are nearly identical to the values for the isotropic case reported in Table \ref{tab:T_H-isotropic}, and that the standard deviations have
the same order of magnitude.\footnote{For large anisotropy $D \lesssim W$ and integer impurity spin $S=1$ and 2, the standard deviation for $\alpha(0.07)$ grows by several orders of magnitude and the obtained average
value is reduced by up to a few percent compared to the isotropic result. The variance mainly increases for magnetic fields smaller than the first effective level crossing field.} In combination with the previous results for
$D \leq 0$, we hence conclude that the effect of the uniaxial anisotropy on the value of $\alpha$ must be very small as long as $|D| \ll W$.

An analysis of the magnetization curves for hard axis anisotropy furthermore shows that the values of the rescaling factor $\eta(0.07, 1)$ (approximately 1.038 for $S=1$, 1.0375 for $S=3/2$, and 1.037 for $S=2$) and the
corresponding proportionality factor $\alpha(0.07)$ are remarkably similar. In case of $S=1$, we have also studied the difference between a local and bulk magnetic field for the two other coupling strengths previously
considered (i.e., for $\rho J =0.05,0.09$). The rescaling factors obtained for $D \ll W$ ($\eta(0.05, 1) \approx 1.0265$ and $\eta(0.09, 1) \approx 1.0497$) are again in remarkable agreement with the corresponding values
of $\alpha(\rho J)$. Moreover, we find that Eqs. (\ref{eq:MgRelation1}) and (\ref{eq:MgRelation2}), with the values of $\eta(\rho J,1)$ as determined for $D>0$, are also suitable to describe the relation between the impurity
magnetization curves for a local and bulk magnetic field for anisotropy $D \leq 0$. As $\mathcal{M}(B,T \approx 0)$ lacks sharp features in the isotropic and easy axis case, the effect of a non-zero electron g-factor is
more subtle, though.

The numerical results thus strongly suggest that the rescaling factor $\eta(\rho J,1)$ and the proportionality factor $\alpha(\rho J)$ take the same value. Furthermore, we find our results to be compatible with the
conclusion of Ref. \onlinecite{Lowenstein1984} that, in case of the isotropic Kondo model with $S=1/2$, the g-factor of the electrons (or equivalently their magnetic moment) is irrelevant for impurity properties in
the limit of infinite bandwidth, corresponding to $\rho J \rightarrow 0$.


In the case of impurity spin $S=1/2$ and vanishing anisotropy $D$, it is also possible to compare the obtained values for $\eta(\rho J,1)$ and $\alpha(\rho J)$ with previously published results. For the
exchange-anisotropic multichannel $S=1/2$-Kondo model with transverse coupling strength $\rho J_{\perp} \ll 1$ (see Eq. (\ref{eq:HEffective1}) for the meaning of the symbols $J_{\perp}$ and $J_{\parallel}$),
it is known that the impurity contributions to the free energy $F_{\text{imp}}(B,T,g_S,g_e)$ for local and bulk magnetic field have the following relation:\cite{Zarand2002}

\begin{eqnarray}
	\label{eq:FImpRelation1}
	F_{\text{imp}}(B,T,g_S,g_S) & = & F_{\text{imp}}(\bar{B},T,g_S,0) \; \; , \\
	\label{eq:FImpRelation2}
	\bar{B} & = & (1-2f\delta/\pi) B \; \; .
\end{eqnarray}

\noindent Here, $f$ is the number of electron channels and $\delta$ is the phase shift generated by the longitudinal coupling $J_{\parallel}$. This result for $F_{\text{imp}}$ is a generalization of the conclusion
that the impurity contribution to the susceptibility $\chi_{\text{imp}}(g_S,g_e)$ \emph{at zero magnetic field} satisfies $\chi_{\text{imp}}(g_S,g_e) = \lambda(g_e/g_S, \delta) \chi_{\text{imp}}(g_S,g_S)$, with a
certain factor $\lambda$, for the single-channel exchange-anisotropic $S=1/2$-Kondo model with $\rho J_{\perp} \ll 1$. \cite{Vigman1978} Using the definition for the impurity contribution to the magnetization,
$M_{\text{imp}} = -\partial F_{\text{imp}}/\partial B$, and the equivalence of $M_{\text{imp}}$ and the impurity magnetization $\mathcal{M}$ for $g_e=0$, the following relation is obtained from Eqs.
(\ref{eq:FImpRelation1}) and (\ref{eq:FImpRelation2}):

\begin{multline}
	\label{eq:MPrediction}	
	M_{\text{imp}}( B,T, g_S, g_S ) = \\
	(1-2f \delta/\pi) \; \mathcal{M}((1-2f \delta/\pi)B,T, g_S, 0) \; .
\end{multline}

On the other hand, the proportionality $\mathcal{M}(B,T \approx 0,g_S,g_S) = \alpha(\rho J) M_{\text{imp}}(B, T \approx 0, g_S, g_S)$ implied by our NRG results can be combined with Eqs. (\ref{eq:MgRelation1})
and (\ref{eq:MgRelation2}) to give:

\begin{multline}
	\label{eq:MProp}
	M_{\text{imp}}(B,T \approx 0, g_S, g_S) = \\
	\frac{1}{\alpha(\rho J)} \; \mathcal{M}\left( \frac{B}{\eta(\rho J,1)}, T \approx 0, g_S, 0 \right) .
\end{multline}

With $f=1$ and the phase shift for the case of an electron band of width $2W$ with constant DOS $\rho=1/2W$,\cite{Costi1999a} $\delta(\rho J_{\parallel}) = \arctan{(\pi \rho J_{\parallel}/4)}$ (note the sign change
with respect to Ref. \onlinecite{Costi1999a}), we compare Eqs. (\ref{eq:MPrediction}) and (\ref{eq:MProp}) and deduce for $\rho J_{\perp} \ll 1$:

\begin{equation}
	\label{eq:factorPrediction}
	\alpha(\rho J_{\parallel}) = \eta(\rho J_{\parallel}, 1) = \frac{1}{1-\frac{2}{\pi}\arctan{(\pi \rho J_{\parallel}/4)}} \; \; .
\end{equation}

\begin{figure}
\centering{\includegraphics[width=1.0 \linewidth]{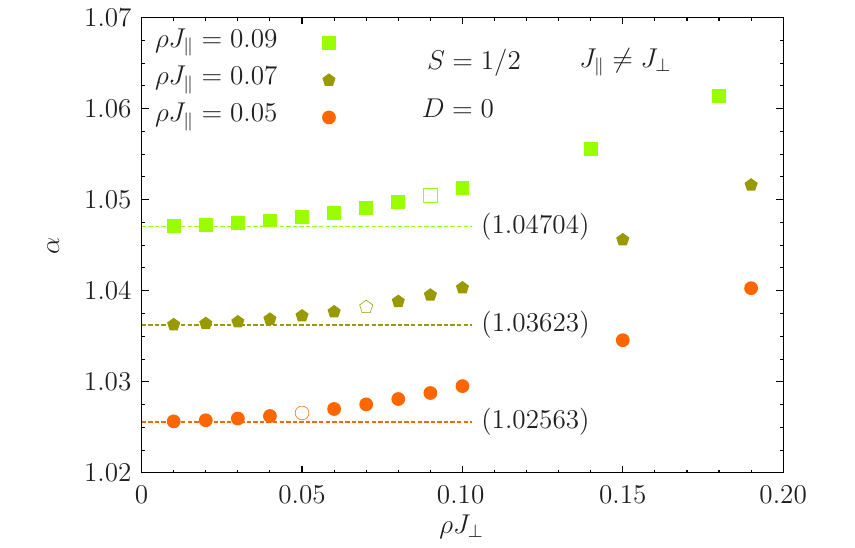}}
%
%
\caption{\label{fig:alpha_vs_rhoJp}(Color online) Proportionality factor $\alpha$ in $\mathcal{M} = \alpha M_{\text{imp}}$ as function of the transverse coupling strength $\rho J_{\perp}$ for impurity spin $S=1/2$,
vanishing uniaxial anisotropy $D$, and three values of the longitudinal coupling $\rho J_{\parallel}$. Dashed horizontal lines mark the values of $\alpha$, given by the numbers in parentheses, that are predicted by
Eq. (\ref{eq:factorPrediction}). Open symbols indicate the proportionality factors for the isotropic case (i.e., $J_{\parallel} = J_{\perp}$), which are also found in Table \ref{tab:T_H-isotropic}. As before, $\alpha$ has
been averaged over magnetic fields $g_S \mu_B B/W \in [10^{-13}, 10^{-1}]$ for $k_BT/W \approx 1.54 \cdot 10^{-15} \approx 0$. The corresponding standard deviations would amount to error bars smaller than
the symbol size.}
\end{figure}

\noindent This equation predicts, in particular, that both the proportionality factor and the rescaling factor tend to 1 in the limit $\rho J_{\parallel} \rightarrow 0$. From Eq. (\ref{eq:factorPrediction}) the following values
for $\alpha(\rho J_{\parallel})$ are obtained: $\alpha = 1.02563$ $(\rho J_{\parallel} = 0.05)$, $1.03623$ $(0.07)$, and $1.04704$ $(0.09)$. In Fig. \ref{fig:alpha_vs_rhoJp}, we present NRG results for $\alpha$ for
an impurity spin $S=1/2$ with vanishing uniaxial anisotropy, but with additional exchange anisotropy. It is seen that the calculated proportionality factors indeed approach the predictions of Eq. (\ref{eq:factorPrediction})
for decreasing transverse coupling strength $\rho J_{\perp}$. For $\rho J_{\perp} = 0.01$ and all three considered values of $\rho J_{\parallel}$, the relative deviation is about $4 \cdot 10^{-5}$. In the isotropic case
(i.e., $J_{\parallel} = J_{\perp} = J$), the relative deviation is less than half a percent, with better agreement for smaller coupling $J$.

Since Eqs. (\ref{eq:FImpRelation1}) and (\ref{eq:FImpRelation2}) also hold for non-zero temperature, the connection between $\mathcal{M}(B, T > 0)$ and $M_{\text{imp}}(B,T > 0)$ has been studied for isotropic
impurity spin $S=1/2,1$, and 3/2, and isotropic coupling $\rho J=0.05,0.07,$ and 0.09. It turns out that $\mathcal{M}$ and $M_{\text{imp}}$ are still proportional for non-zero temperature and that, for
$g_S \mu_B B / W \in [10^{-13}, 10^{-1}]$, the relative deviation between the proportionality factor and the corresponding value $\alpha(\rho J, T \approx 0)$ is less than 1 $\text{\textperthousand}$ for thermal
energies $k_BT/W \leq 10^{-2}$.

\subsection{\label{subsec:fieldInducedKondoEffect}Field-induced Kondo effect}

To better understand the steps in the low temperature magnetization curves for large anisotropy $D>0$ (cf. Figs. \ref{fig:M_vs_BD-hardAxis} and \ref{fig:FIKE_vs_B} a), we now derive effective (simplified) models
near groundstate level crossings (LCs) of the corresponding free anisotropic spin (see Fig. \ref{fig:M_vs_B-hardAxis-freeSpin}). These models are approximations to the full Hamiltonian in the limit $D \rightarrow \infty$.

\subsubsection{\label{subsubsec:FIKE_effectiveModel}Effective models near groundstate level crossings in the limit of arbitrarily large anisotropy}

For given impurity spin $S \geq 1$ let us consider one of the groundstate level crossings of the corresponding free anisotropic spin (cf. Fig. \ref{fig:M_vs_B-hardAxis-freeSpin}). If $D$ is large, then the two levels which
cross in the groundstate are energetically well separated from the rest of the spectrum in the vicinity of the LC field. As an approximation for the full impurity model \emph{near this free LC field}, we therefore project the
impurity degrees of freedom in Hamiltonian (\ref{eq:HTotal}) onto the two impurity states involved in the free LC. This way, the impurity Hilbert space is reduced to two states and the impurity spin $S$ can thus be mapped
to an effective spin-1/2. While the projection becomes exact only in the limit $D \rightarrow \infty$, we expect it to be a quantitative approximation for $D \gg W$ and $D \gg k_B T$.

The mapping of the impurity spin to a spin-1/2 is an extension of the ideas from Refs. \onlinecite{Schiller2008} and \onlinecite{Zitko2008b}. In contrast to the case of zero magnetic field that has been studied there, we
do not project onto impurity doublets with $M=\pm1/2$ (see below). Furthermore, at each LC, i.e., for each step in the magnetization curve, the impurity is reduced to a different pair of states. As a consequence, different
parameters of the effective model are obtained at each LC.

We intend to use the effective models to determine the magnetic fields at which steps appear in the impurity magnetization curves for large anisotropy $D$ (cf. Fig. \ref{fig:M_vs_BD-hardAxis}), and to investigate how the
properties of the full impurity model differ near the various free LCs (as indicated, e.g., by Fig. \ref{fig:FIKE_vs_B} a). Compared to the full model, the effective models are numerically less demanding as they feature an
impurity spin-1/2 independent of the value of $S$, and they allow to study the effect of the different terms appearing in the effective Hamiltonian. 

To be specific, we consider the two impurity states with magnetic quantum numbers $-M$ and $-(M+1)$ (assuming $M \geq 0$), which cross at the free LC field $B_{M}=(2M+1)D/g_S\mu_B$, and project
Hamiltonian (\ref{eq:HTotal}) onto them. The effective model is determined by requiring that its matrix representation be equal to that of the full model in the chosen subspace. Note that we have to introduce
new impurity states by shifting the magnetic quantum number in order to map the impurity spin to a spin-1/2. This mapping then corresponds to the following replacements for the impurity spin operators:

\begin{eqnarray}
	\label{eq:mappingx} \op{S}^x \; & \rightarrow & \; \sqrt{(S-M)(S+M+1)} \; \op{s}^x \; , \\
	\label{eq:mappingy} \op{S}^y \; & \rightarrow & \; \sqrt{(S-M)(S+M+1)} \; \op{s}^y \; , \\
	\label{eq:mappingz} \op{S}^z \; & \rightarrow & \; \op{s}^z - (M+1/2) \; \op{\mathbbm{1}}_s \; .
\end{eqnarray}

\noindent The impurity now has spin $s=1/2$ and the replacement for $(\op{S}^z)^2$ directly follows from that for $\op{S}^z$. In the parameter regime in which the projection is valid, the mapping
for $\op{S}^z$ leads to the following connection between the impurity magnetization of the full and effective model:

\begin{equation}
	\label{eq:MRelation}
	\mathcal{M}/g_S \mu_B = - \langle \op{S}^z \rangle \approx - \langle \op{s}^z \rangle + (M+1/2) \; .
\end{equation}

\noindent There is an analogous relationship for the impurity contribution to the magnetization $M_{\text{imp}}$. According to its definition, the impurity contribution to the magnetic susceptibility $\chi_{\text{imp}}$ is not
affected by the shift of the magnetic quantum numbers.

Applying mappings (\ref{eq:mappingx}) to (\ref{eq:mappingz}) to the full impurity model (\ref{eq:HTotal}) and dropping all constant terms, the following Hamiltonian is obtained:

\begin{eqnarray}
	\label{eq:HEffective1} \op{H}_s(S,M) & = & \sum_{\bm{k},\sigma}{\varepsilon_{\bm{k} \sigma} \opDag{c}_{\bm{k} \sigma} \op{c}_{\bm{k} \sigma}} - \kappa \op{s}^z_0 \\
	\nonumber & + & J_{\perp} ( \op{s}_0^x \op{s}^x + \op{s}_0^y \op{s}^y ) + J_{\parallel} \op{s}_0^z \op{s}^z \\
	\nonumber & + & g_S \mu_B (B - B_M) \op{s}^z \; ,
\end{eqnarray}

\noindent with the set of parameters

\begin{eqnarray}
	\label{eq:effPara1} \varepsilon_{\bm{k} \sigma} & = & \varepsilon_{\bm{k}} + \sigma g_e \mu_B B \\
					&& (\sigma = \pm 1/2) \; , \nonumber \\
	\label{eq:effPara2} \kappa & = & (M+1/2) \, J \; , \\
	\label{eq:effPara3} J_{\perp} & = & \sqrt{(S-M)(S+M+1)} \, J \; , \\
	\label{eq:effPara4} J_{\parallel} & = & J \; , \\
	\label{eq:effPara5} B_M & = & (2M+1)D/g_S \mu_B \; .
\end{eqnarray}

\noindent In contrast to the full Hamiltonian, $\op{H}_s(S,M)$ is exchange anisotropic with $J_{\perp} > J_{\parallel}$: $J_{\perp} / J_{\parallel}$ grows with $S$ and decreases with increasing $M$ (while always
present, the exchange anisotropy thus becomes weaker with every further LC). The Zeeman term for the impurity is now expressed relative to the free LC field $B_M$. $\op{H}_s(S,M)$ furthermore contains the new
term $-\kappa \op{s}^z_0$ representing an effective magnetic field, which couples to the electron spin at the origin and points in the opposite direction of the external field $B$. With respect to NRG, this term can be
regarded as spin-dependent scattering at the zeroth site of the Wilson chain. It breaks the invariance under a spinflip transformation ($\op{c}_{\bm{k}\sigma} \rightarrow \op{c}_{\bm{k}-\sigma}$ and
$\opVec{s} \rightarrow (\op{s}^x, -\op{s}^y, -\op{s}^z$)), which $\op{H}_s(S,M)$ would otherwise possess for $B=B_M$. While the scattering parameter $\kappa$ grows with $M$, the ratio $\kappa/B_M$ is
independent of $M$. Starting with the second LC, $\kappa$ is larger than $J_{\parallel}$. The ratio $\kappa/J_{\perp}$, which at first is smaller than 1, also grows with $M$ and eventually becomes greater than 1
if $S$ is large enough.

As an analogue to the free LC field $B_M$, we call the magnetic field $B_{\text{ELC}}$ for which the impurity magnetization vanishes at zero temperature (i.e., the two impurity levels are effectively degenerate),
\emph{``effective level crossing (ELC) field''}:

\begin{equation}
	\langle \op{s}^z \rangle(B_{\text{ELC}},T=0)=0 \; .
\end{equation}

\noindent In the parameter regime in which the mapping to a spin-1/2 is valid, there is a step in the impurity
magnetization curve of the full model at the ELC and, according to Eq. (\ref{eq:MRelation}), the value of $\mathcal{M}$ at the ELC is $\mathcal{M} \approx g_S \mu_B (M+1/2)$. In the following, we discuss the
properties of Hamiltonian (\ref{eq:HEffective1}) in more detail for the two different cases $g_e>0$ and $g_e=0$.

Let us begin with the case $g_e>0$. As the free LC field $B_M$ is proportional to $D$, the limit $D \rightarrow \infty$ also corresponds to the limit $B \rightarrow \infty$. A non-zero Zeeman coupling of the electrons
therefore leads to their complete polarization so that formally they may be replaced with spinless fermions (corresponding to spin-down electrons). Since the remaining fermion band is then completely filled, all
interaction terms vanish and the electrons can be completely eliminated from the problem. For $g_e>0$ and arbitrarily large $D$, Hamiltonian (\ref{eq:HEffective1}) thus reduces to a pure spin model:

\begin{equation}
	\label{eq:HEffective2}
	\op{H}_{\text{eff}}^{(g_e>0)}(\widetilde{B}) = g_S \mu_B \left( \widetilde{B} - \frac{J/2}{g_S\mu_B} \right) \op{s}^z \; .
\end{equation} 

\noindent Here, we have introduced a relative magnetic field $\widetilde{B} = B - B_M$. As the only remnant of the interaction between impurity and electrons, a shift of the free LC field remains. This shift is
positive for antiferromagnetic coupling $J>0$ and only depends on the coupling strength, but not on $S$ or $M$. It is thus the same for all LCs. From the effective model (\ref{eq:HEffective2}) we learn that the
ELC fields eventually exceed the free LC fields for $g_e>0$ and large anisotropy $D$ (cf. Fig. \ref{fig:M_vs_BD-hardAxis}).

We now turn to the case of a local magnetic field. Setting $g_e = 0$ and using the relative field $\widetilde{B}$, Hamiltonian (\ref{eq:HEffective1}) becomes the effective model for arbitrarily large $D$:

\begin{equation}
	\label{eq:HEffective3}
	\op{H}_{\text{eff}}^{(g_e=0)}(\widetilde{B};S,M) = \left. \op{H}_s(S,M) \right|_{g_e = 0} \; .
\end{equation}

\noindent We are particularly interested in the properties of $H_{\text{eff}}^{(g_e=0)}(\widetilde{B};S,M)$ at the ELC field $\widetilde{B}_{\text{ELC}} = B_{\text{ELC}} - B_M$. Due to the scattering term, the effective
model does not exhibit a spinflip-invariance at the ELC. It therefore seems that the ELC is not characterized by special symmetry properties. A spin-\emph{independent} (potential) scattering term can be treated by
transforming to scattering states which diagonalize the electronic part of the Hamiltonian (cf. App. C of Ref. \onlinecite{Krishnamurthy1980b}). Although such a transformation can be easily adapted to the case of
spin-\emph{dependent} scattering, it does not seem to yield the intended results. The approximation which is used in the spin-independent case (a modification of the density of states at the Fermi level expressed
by an effective coupling parameter) \cite{Krishnamurthy1980b} would restore spinflip-invariance for $\widetilde{B}=0$ in the spin-dependent case and would thus erroneously imply
$\langle \op{s}^z \rangle(\widetilde{B}=0,T)=0$. Instead, we are going to use NRG to determine the ELC field $\widetilde{B}_{\text{ELC}}$ and to study the properties of
$H_{\text{eff}}^{(g_e=0)}(\widetilde{B}_{\text{ELC}}; S,M)$.

For the interpretation of the properties of the full Hamiltonian (\ref{eq:HTotal}) near the ELCs, we are going to use the main results for the effective model $\op{H}_{\text{eff}}^{(g_e=0)}(\widetilde{B}_{\text{ELC}};S,M)$.
These are explicitly demonstrated in Sec. \ref{subsubsec:FIKE_effectiveModelProperties} and are summarized in the following. At an ELC the impurity spin $s=1/2$ of the effective model is Kondo screened for
$T \rightarrow 0$. The temperature dependence of the impurity contribution to the entropy at an ELC is described by the corresponding universal function for the isotropic $S=1/2$-Kondo model with zero magnetic
field. Since the parameters of the effective model are different near each free LC (see Eqs. (\ref{eq:effPara2}) and (\ref{eq:effPara3})), there is also a different Kondo temperature $T_K$ at each ELC. It turns out that
$T_K$ decreases with increasing $M$, i.e., $T_K$ becomes smaller with every further ELC.

\subsubsection{\label{subsubsec:FIKE_BDependence}Magnetic field dependence of impurity contributions near effective level crossing fields}

\begin{figure}
\centering{\includegraphics[width=1.0 \linewidth]{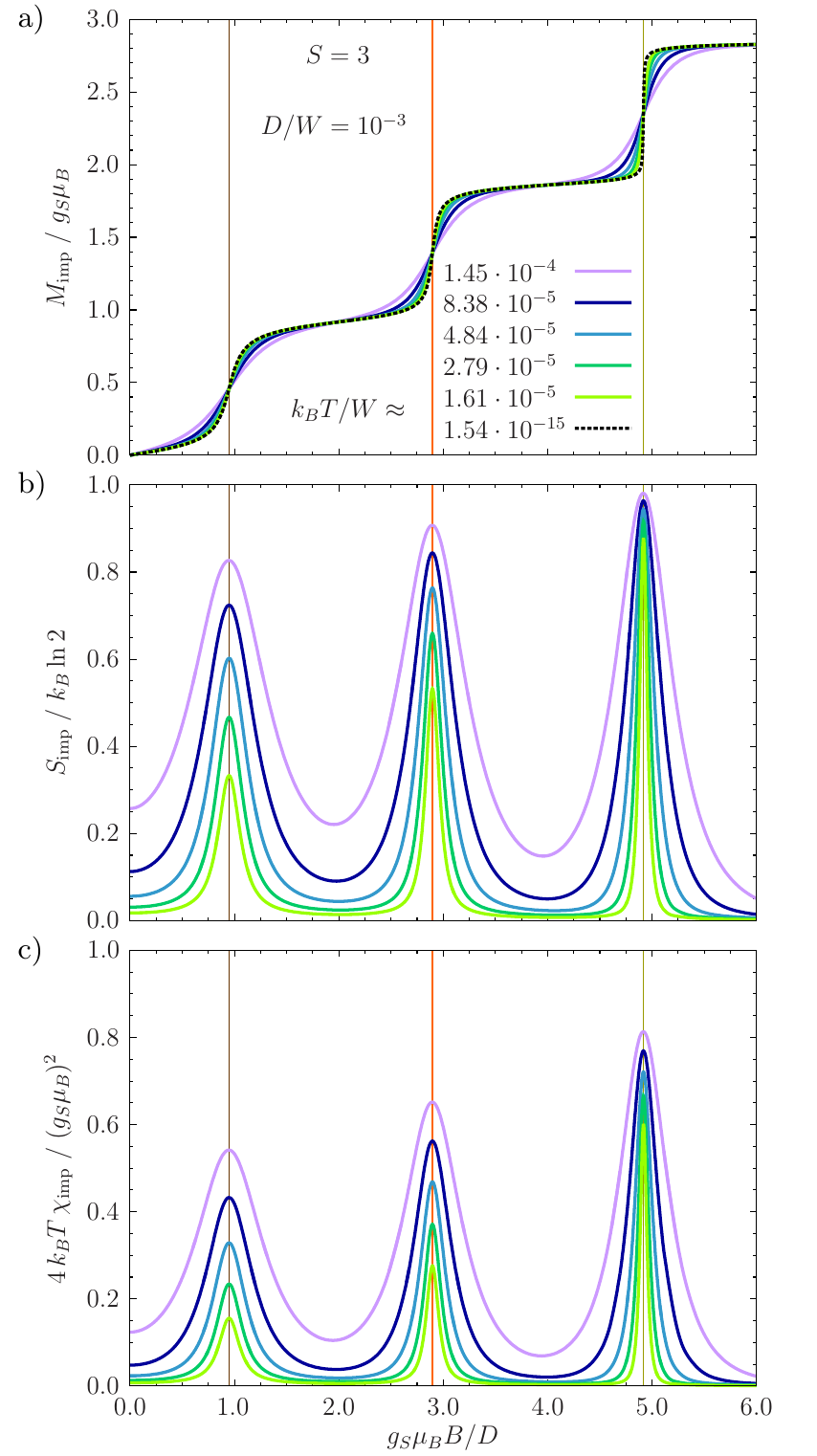}}
%
%
%
%
\caption{\label{fig:FIKE_vs_B}(Color online) Impurity contribution to \textbf{a)} the magnetization $M_{\text{imp}}$, \textbf{b)} the entropy $S_{\text{imp}}$, and \textbf{c)} the effective moment $T \chi_{\text{imp}}$
as function of magnetic field for impurity spin $S=3$, hard axis anisotropy $D/W = 10^{-3}$, and several temperature values. In plot a, the dashed curve shows $M_{\text{imp}}$ for $k_B T / W \approx 0$, and
for the curves in plots b and c temperature increases from bottom to top. Thin vertical lines indicate the (equal) peak positions for $S_{\text{imp}}$ and $T \chi_{\text{imp}}$. As before, the coupling strength is
chosen as $\rho J = 0.07$ and equal g-factors are used.}
\end{figure}

Armed with the effective model $\op{H}_{\text{eff}}^{(g_e=0)}(\widetilde{B};S,M)$ for a local magnetic field, we now study in detail the field dependence of typical impurity contributions of the full impurity model for
moderately large anisotropy $D>0$. In Fig. \ref{fig:FIKE_vs_B} results for $M_{\text{imp}}(B)$, $S_{\text{imp}}(B)$, and $T\chi_{\text{imp}}(B)$ are shown for impurity spin $S=3$ and anisotropy $D/W=10^{-3}$. As
before, equal g-factors have been assumed. \footnote{It seems permissible to use the effective model for $g_e=0$ as a \emph{guide} for the interpretation of the results in Fig. \ref{fig:FIKE_vs_B} since we are primarily
interested in the behavior of the impurity magnetization $\mathcal{M}$ whose dependence on $g_e$ we know. As demonstrated earlier for not too large $D$, $\mathcal{M}$ is proportional to $M_{\text{imp}}$ for
equal g-factors and a reduction of $g_e$ effectively rescales the magnetic field argument of $\mathcal{M}$. This of course changes the ELC field, but preserves the shape of the impurity magnetization curve (cf. Fig.
\ref{fig:M_vs_B-hardAxis-S1-gRatio}).}

Let us start with a discussion of the magnetization curves depicted in Fig. \ref{fig:FIKE_vs_B} a. According to the previously considered behavior of the impurity magnetization $\mathcal{M}$ and its connection with
$M_{\text{imp}}$, there are also steps in $M_{\text{imp}}(B)$ at low temperature. These steps have finite widths for $T \rightarrow 0$ and are smeared out for sufficiently high temperature. It is noticeable that the steps
have different widths: those occurring at larger magnetic field are steeper. Fig. \ref{fig:FIKE_vs_B} a furthermore indicates that the influence of non-zero temperature is different for the different steps. In contrast, for a
free anisotropic spin the steps in the magnetization become discontinuous for $T \rightarrow 0$ and the effect of non-zero (small) temperature is the same for all of them (see Fig. \ref{fig:M_vs_B-hardAxis-freeSpin}). 
In the chosen representation of Fig. \ref{fig:FIKE_vs_B} a, the pseudo-plateaus between the steps become flatter in the direction of increasing magnetic field and approach the true plateaus of the free anisotropic spin
from below for growing anisotropy $D$.

The behavior of $M_{\text{imp}}(B)$ as shown in Fig. \ref{fig:FIKE_vs_B} a can be understood by considering the magnetic field dependence of $M_{\text{imp}}$ for the isotropic $S=1/2$-Kondo model with
$g_e=g_S$. In this case, $M_{\text{imp}}(B,T=0)$ is described by a universal function $f(x)$ with the variable $x = g_S \mu_B B/k_B T_H$ (and $T_H \propto T_K$, cf. Sec. \ref{subsec:M_vs_B-isotrop}).
\cite{Tsvelick1983} $f(x)$ is linear in $x$ for $x \ll 1$ and thus the slope of $M_{\text{imp}}(B)$ for small fields is higher if the Kondo temperature is smaller. \cite{Tsvelick1983,Andrei1983} This relation is
also expressed by Wilson's definition of the Kondo temperature: \cite{Wilson1975} The zero-field susceptibility for $T \ll T_K$ is a $T_K$-dependent constant: $\chi_{\text{imp}}(B=0,T \ll T_K) = \text{const.}/T_K$.
Combined with the prediction of the effective model with $g_e=0$ for the Kondo temperatures at the different ELCs, this explains why the steps in the magnetization curve of the full model have different widths
at zero temperature. In case of the $S=1/2$-Kondo model, temperature has to reach the scale of $T_K$ to become relevant for the zero-field susceptibility. \cite{Wilson1975} For this reason, thermal broadening
of a step in $M_{\text{imp}}(B)$ begins at lower temperature if the step occurs at a later ELC with smaller $T_K$. Furthermore, away from an ELC the magnetization reaches values of the order of the respective
saturation value for smaller magnetic fields (relative to the ELC field) if the Kondo temperature at the ELC is lower. It subsequently enters the regime of very slow growth towards saturation, which shows up in
Fig. \ref{fig:FIKE_vs_B} a in the form of a pseudo-plateau. This also explains why pseudo-plateaus between later ELCs with smaller $T_K$ are flatter.

We now continue with a discussion of the magnetic field dependence of the impurity contribution to the entropy and the effective moment. Results for $S_{\text{imp}}(B)$ and $T\chi_{\text{imp}}(B)$ at low temperature
$k_B T < D$ are shown in Fig. \ref{fig:FIKE_vs_B} b and c, respectively. In both cases we observe peaks of varying height and width whose positions coincide with those of the steps in $M_{\text{imp}}(B)$. We find
that the peaks become both higher and narrower with every further ELC. If the temperature is reduced, the peak heights decline and at the same time, if $T$ is not too low, the peaks become sharper. It is noticeable
that there is a temperature below which the width of the first peak in both $S_{\text{imp}}$ and $T\chi_{\text{imp}}$ varies only little as a function of $T$.

At zero temperature both $S_{\text{imp}}$ and $T\chi_{\text{imp}}$ vanish for all magnetic fields. In case of the entropy, the reason is that the magnetic field either leads to a non-degenerate groundstate or it creates
an effective impurity doublet which is then Kondo screened. The effective moment, on the other hand, has to go to zero since the slope of $M_{\text{imp}}(B)$ at zero temperature, i.e., $\chi_{\text{imp}}(B,T=0)$, is
finite for all fields. For large anisotropy $D$, the temperature dependence of the peak heights is determined by the pseudo-spin-1/2 Kondo effects that take place at the ELCs (see the next section and Fig.
\ref{fig:FIKE_vs_T} for details). By recollecting results for the $S=1/2$-Kondo model, \cite{Sacramento1989} we can furthermore understand the different peak widths and their temperature dependence. In case of
the $S=1/2$-Kondo effect, the Zeeman energy $g_S \mu_B B$ has to reach the energy scale of $\max{(k_BT,k_BT_K)}$ in order to considerably suppress $S_{\text{imp}}$ and $T\chi_{\text{imp}}$.
\cite{Sacramento1989} In particular, since the lowest temperatures considered in Fig. \ref{fig:FIKE_vs_B} are smaller than $T_K$ at the first ELC (cf. the indicated temperature range in Fig. \ref{fig:FIKE_vs_T} a),
temperatures are reached for which the thermal broadening of the first peak is small.

\subsubsection{\label{subsubsec:FIKE_TDependence}Temperature dependence of impurity contributions at effective level crossing fields}

\begin{figure}
\centering{\includegraphics[width=1.0 \linewidth]{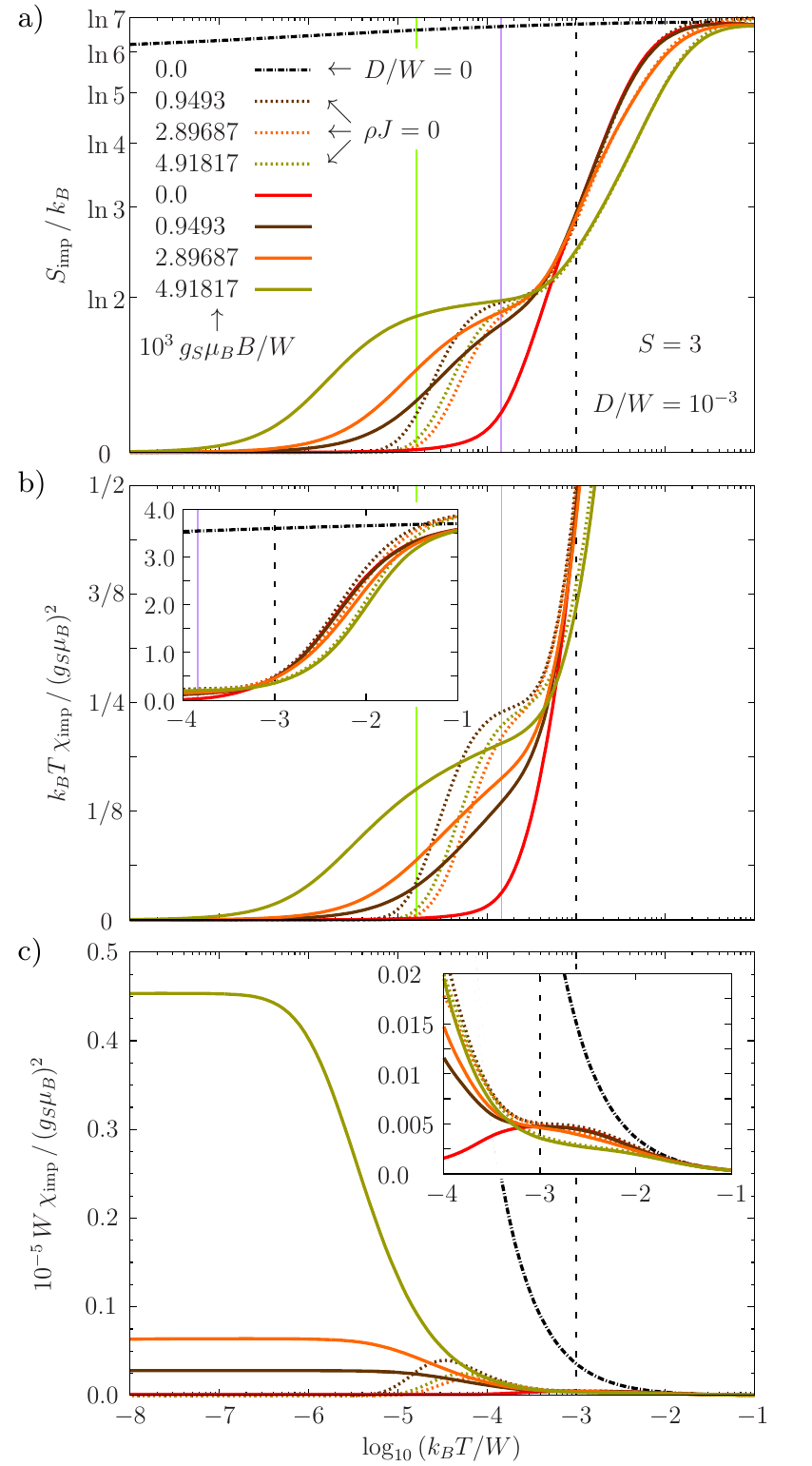}}
%
%
%
%
\caption{\label{fig:FIKE_vs_T}(Color online) Impurity contribution to \textbf{a)} the entropy $S_{\text{imp}}$, \textbf{b)} the effective moment $T \chi_{\text{imp}}$, and \textbf{c)} the magnetic
susceptibility $\chi_{\text{imp}}$ as function of temperature for impurity spin $S=3$ and hard axis anisotropy $D/W = 10^{-3}$. The chosen non-zero magnetic fields correspond to the peak
positions for $S_{\text{imp}}$ and $T \chi_{\text{imp}}$ according to Fig. \ref{fig:FIKE_vs_B}. Also shown are NRG results for vanishing coupling strength (dashed lines) and for the isotropic case
(dash-dotted lines). Solid vertical lines delimit the temperature interval considered in Fig. \ref{fig:FIKE_vs_B}, whereas dashed vertical lines indicate the thermal energy for which $k_BT=D$. In
the insets, data from the main plots is presented for a reduced temperature range.}
\end{figure}

Fig. \ref{fig:FIKE_vs_T} shows the temperature dependence of $S_{\text{imp}}$, $T\chi_{\text{imp}}$, and $\chi_{\text{imp}}$ at the three ELCs that occur for the example considered in Fig. \ref{fig:FIKE_vs_B}.
In case of entropy and effective moment it is thus demonstrated how the peak heights in Figs. \ref{fig:FIKE_vs_B} b and c decrease when the temperature is lowered.

Let us begin with a discussion of the results for $S_{\text{imp}}(T)$ shown in Fig. \ref{fig:FIKE_vs_T} a. $S_{\text{imp}}(B=0,T)$ for the isotropic impurity with $S=3$ interpolates between the limiting values
of $k_B \ln{7}$ (for $T \rightarrow \infty$) and $k_B \ln{6}$ (for $T=0$), according to the screening of half a magnetic moment. In contrast, for the anisotropic impurity in zero field all higher-lying impurity levels are frozen
out on the temperature scale $k_BT \approx D$ so that $S_{\text{imp}}$ quickly drops to zero. \cite{Zitko2008b} At high temperatures $k_B T \gg D$, on the other hand, the anisotropic impurity behaves more and
more like an isotropic impurity. For the three ELC fields, we observe qualitatively different behavior at low temperature $k_BT < D$ compared to the case of zero field: Starting at $S_{\text{imp}} \approx k_B \ln{2}$,
the effective impurity doublet, which is formed due to the magnetic field, undergoes Kondo screening with a value of $T_K$ that decreases with every further ELC. While there is fair agreement with the results for
a decoupled impurity at $k_BT >D$, low temperature $k_BT \ll D$ reveals that the ELC fields are not equal to the free LC fields. For this reason, all states except the respective non-degenerate groundstate are
ultimately frozen out for $T \rightarrow 0$.

We continue with a discussion of the temperature dependence of the effective moment $T\chi_{\text{imp}}$ as illustrated in Fig. \ref{fig:FIKE_vs_T} b. For the isotropic impurity with $S=3$,
$k_BT\chi_{\text{imp}}(B=0,T)/(g_S\mu_B)^2$ obeys a Curie law at both high and low temperature and goes from $S(S+1)/3=4$ (for $T \rightarrow \infty$) to $(S-1/2)(S+1/2)/3=35/12$
(for $T=0$). For the anisotropic impurity in zero magnetic field, on the other hand, the increasing thermal reduction to a non-magnetic groundstate for $k_BT < D$ leads to a vanishing
effective moment at zero temperature. \cite{Zitko2008b} At very high temperature $k_BT \lesssim W$ (not shown), we find the expected agreement between $T\chi_{\text{imp}}(B=0)$
for the isotropic and anisotropic case. The effective moment of the decoupled anisotropic impurity first goes to about 1/4 at $k_BT \lesssim D$ (according to the susceptibility of a doublet
with $\Delta M=1$), but then quickly drops to zero because of the non-degenerate groundstate with good magnetic quantum number. For the anisotropic impurity at the ELC fields, we
again observe Kondo screening for $T \rightarrow 0$ with different $T_K$, starting with an effective moment of $k_BT \chi_{\text{imp}}/(g_S\mu_B)^2 \approx 1/4$ at $k_B T \lesssim D$.

Finally, we take a look at the results for the impurity contribution to the susceptibility $\chi_{\text{imp}}(T)$ that are presented in Fig. \ref{fig:FIKE_vs_T} c. According to its definition, $\chi_{\text{imp}}$ is
just the slope of $M_{\text{imp}}$ and thus directly yields information about the width of the steps in the magnetization curve shown in Fig. \ref{fig:FIKE_vs_B} a. For an isotropic impurity with $S \geq 1$,
the zero-field susceptibility at low temperature is described by a Curie law and thus diverges for $T \rightarrow 0$. In contrast, $\chi_{\text{imp}}(B=0,T)$ for the anisotropic impurity has a maximum at
$k_BT \approx D$ and vanishes for zero temperature. Since the ELCs lie close to the free LCs in this example, the susceptibility for the decoupled anisotropic impurity displays a maximum at
a thermal energy of the order of the level splitting before falling off at low temperature. At the ELCs, $\chi_{\text{imp}}(T)$ for the anisotropic impurity saturates at a finite value for $T \rightarrow 0$ that
increases with decreasing Kondo temperature.

Recently, it has been demonstrated that a field-induced Kondo effect also occurs for Hamiltonian (\ref{eq:HTotal}) with easy axis anisotropy $D<0$, additional transverse anisotropy $E$, and a local
magnetic field aligned along the $x$-axis. \cite{Zitko2010c}

\subsubsection{\label{subsubsec:FIKE_effectiveModelProperties}Properties of the effective model for vanishing electron g-factor}

We now return to the effective model for $g_e=0$ given by Hamiltonian (\ref{eq:HEffective3}) in order to study its properties in greater detail. The ELC field $\widetilde{B}_{\text{ELC}}$ and the Kondo
temperature $T_K^{\text{ELC}}$ \emph{at the ELC field} are determined as function of the parameters $J_{\parallel}$, $J_{\perp}$, and $\kappa$ of the effective model or, respectively, as function of the
parameters $J$, $S$, and $M$ of the full Hamiltonian (according to Eqs. (\ref{eq:effPara2}) to (\ref{eq:effPara4})). As a prerequisite, we have to figure out how to reliably extract these quantities from the
NRG results.

To determine $\widetilde{B}_{\text{ELC}}$, the impurity magnetization in units of $g_S \mu_B$ for the effective model, $- \langle \op{s}^z \rangle(\widetilde{B})$, is calculated for low temperature $k_B T \ll W$.
In the vicinity of an ELC, i.e., near its root, the impurity magnetization depends linearly on the (relative) magnetic field $\widetilde{B}$. The root, which corresponds to $\widetilde{B}_{\text{ELC}}$ at $T=0$, can
therefore be determined by performing a linear fit to the numerical data. However, the following complication arises: The position of the root of $\langle \op{s}^z \rangle(\widetilde{B})$ depends on the value
of the twist parameter $z$ and thereby on the discretization of the electron band. On the contrary, a physically meaningful result for the ELC field should display only a weak dependence on the numerical
parameters of an NRG calculation in order to accurately reflect the continuum limit $\Lambda \rightarrow 1$. It turns out that a standard $z$-averaging, i.e., an averaging of the impurity magnetization curves
for different values of $z$ at fixed temperature, is not reasonable at this point. Near an ELC, such an averaging in general introduces artifacts into the averaged curve because of non-linear components which
some of the $z$-dependent curves might already comprise. Similar numerical artifacts are found in the $z$-averaged magnetization curves of the full model for large hard axis anisotropy. Upon closer inspection,
one discovers that $z$-averaging divides the total height of a magnetization step into smaller ``sub-steps'' of equal height whose number corresponds to the number of $z$-values used.


\begin{figure}
\centering{\includegraphics[width=1.0 \linewidth]{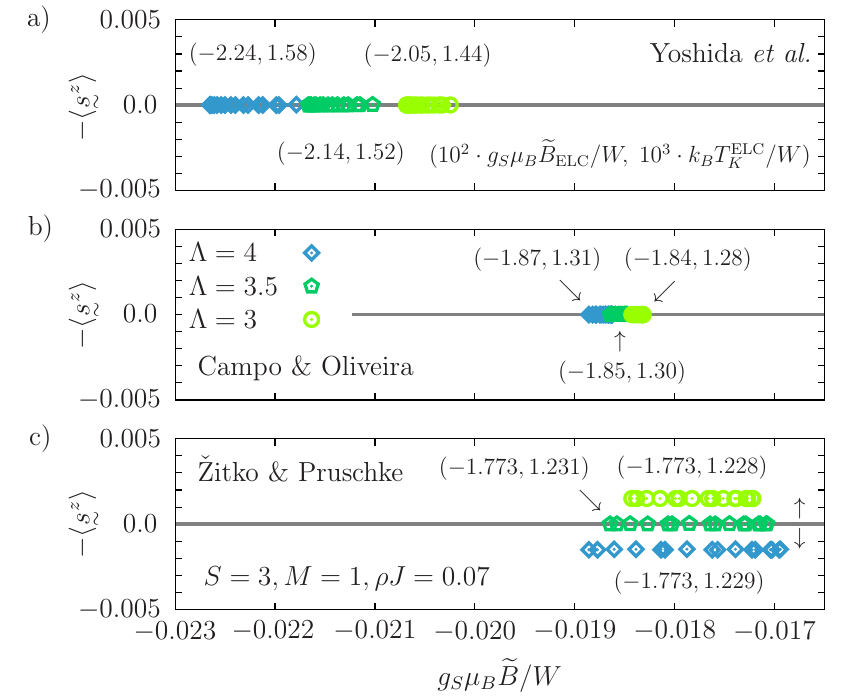}}
%
%
%
%
\caption{\label{fig:ELC-Comparison}(Color online) Impurity magnetization (in units of $g_S\mu_B$) for the effective Hamiltonian with $g_e=0$ and parameters according to Eqs. (\ref{eq:effPara2})
to (\ref{eq:effPara4}) as function of the relative magnetic field for $k_BT/W \approx 10^{-15} \approx 0$. The results have been calculated using the discretization schemes by a) Yoshida \emph{et al.} (with
correction factor \cite{Krishnamurthy1980a, Campo2005,Bulla2008} $A_{\Lambda}$), \cite{Yoshida1990, Oliveira1994, Costa1997} b) Campo and Oliveira, \cite{Campo2005} and c) \v{Z}itko and
Pruschke. \cite{Zitko2009a, Zitko2009b} For each discretization scheme, results are presented for three values of the discretization parameter $\Lambda$ and 16 values of the twist parameter $z$
(i.e., $z_i = i/16$ with $i \in \{ 1,2,...,16\}$). In plot c, data points are vertically offset to enhance legibility. Numbers in parentheses denote the ELC field $\widetilde{B}_{\text{ELC}}$ and the Kondo
temperature $T_K^{\text{ELC}}$ at the ELC field, respectively (cf. main text).}
\end{figure}

For one set of parameters of the effective model, the dependence of the impurity magnetization root on the discretization of the electron band is demonstrated in Fig. \ref{fig:ELC-Comparison}. There, the three
common discretization schemes are compared for three different values of the discretization parameter $\Lambda$ and, \emph{only in this example}, for 16 values of the twist parameter $z$. In all cases, we observe
a spread of the position of the impurity magnetization root with respect to $z$. This variation decreases for smaller values of $\Lambda$ and is always largest when using the discretization scheme by \v{Z}itko
and Pruschke (\v{Z}P). The spread due to $z$ defines a magnetic field interval which, in case of the discretization by Yoshida \emph{et al.} (Y) and Campo and Oliveira (CO), moves towards larger fields when
$\Lambda$ is reduced. In contrast, the \v{Z}P discretization leads to nested intervals so that an interval for smaller $\Lambda$ is wholly contained in an interval for larger $\Lambda$. In the special case $z=1$
(which corresponds to the smallest root), the CO and \v{Z}P discretization give the same result. \cite{Zitko2009a} Note that the $\Lambda$-dependence of the data shown in Fig. \ref{fig:ELC-Comparison} is
consistent with an agreement of the results of all three discretization schemes in the continuum limit $\Lambda \rightarrow 1$. However, in order to obtain reliable information about the continuum limit it would
be necessary to perform an impractical extrapolation in $\Lambda$ when using the Y or CO discretization. On the contrary, the \v{Z}P discretization apparently allows to make a dependable statement about the
limit $\Lambda \rightarrow 1$ on the basis of results for only a single value of the discretization parameter: Fig. \ref{fig:ELC-Comparison} suggests that the continuum value of $\widetilde{B}_{\text{ELC}}$ lies in
the magnetic field interval that is spanned by the $z$-dependent roots for $\Lambda > 1$. It turns out that, using the \v{Z}P discretization, one can obtain a better approximation for the ELC field by averaging over
the $z$-dependent impurity magnetization roots since the resulting mean value displays only a weak dependence on $\Lambda$ (cf. the numbers in parentheses in Fig. \ref{fig:ELC-Comparison}). The spread of
the roots with respect to $z$ then provides a safe error estimate for the mean value (amounting to a relative deviation of about 3 to 4 \% for $\Lambda=3$). However, the dependence of the mean value on $\Lambda$
indicates that such an error estimate is far too pessimistic. For the \v{Z}P discretization with $\Lambda=3$ and \emph{four $z$-values}, we expect that the relative error of the obtained ELC field $\widetilde{B}_{\text{ELC}}$
is about one order of magnitude smaller than suggested by the roots' dependency on $z$.

Having determined $\widetilde{B}_{\text{ELC}}^z$ for all values of $z$, we can study the thermodynamic properties of the effective model at the ELC field by calculating the impurity contribution to the entropy
$S^z_{\text{imp}}(T, \widetilde{B}_{\text{ELC}}^z)$ \emph{for each $z$ at the respective ELC field}. For thermal energies that are small compared to the bandwidth, the NRG results for
$S^z_{\text{imp}}(T, \widetilde{B}_{\text{ELC}}^z)$ can be aligned with the known universal temperature dependence of the entropy for the isotropic Kondo model with $S=1/2$. We find, however, that there is
a dip in the entropy for thermal energies close to the band edge that becomes more pronounced for stronger scattering $\kappa$. The Kondo temperature $T_K^{\text{ELC},z}$ characterizing the universal
behavior of $S^z_{\text{imp}}(T, \widetilde{B}_{\text{ELC}}^z)$ is obtained in the following way for each value of $z$: By comparing with the Bethe Ansatz solution for the impurity contribution to the magnetization
in Sec. \ref{subsec:M_vs_B-isotrop}, the value of $T_K$ (as occurring in the BA solution of the Kondo model) for the isotropic NRG results is known (cf. Table \ref{tab:T_H-isotropic} and Eq. (\ref{eq:THandTK})).
As a first step, the result for $S^z_{\text{imp}}(T, \widetilde{B}_{\text{ELC}}^z)$ is restricted to the linear low-temperature regime in which a continuous curve is produced using a linear fit. This fit then allows to
determine the value of $T_K^{\text{ELC},z}$ by comparing with an isotropic NRG result for $S=1/2$ with known Kondo temperature $T_K^{\text{iso}}$. A better approximation for the Kondo temperature at the
ELC field $T_K^{\text{ELC}}$ is again obtained by averaging over the $z$-dependent values. Regarding the variation with respect to $z$ and the error estimate for the mean value, comparable statements hold
true as in the case of the ELC field.

Before continuing, we note that the root of the impurity magnetization depends on temperature. This is possible since its value is not determined by symmetry properties of the effective model. When the
temperature exceeds the Kondo temperature at the ELC field, two effects eventually occur: The slope of the magnetization curve decreases and the root moves towards larger relative magnetic fields.


We now study how the parameter $\kappa$, which corresponds to spin-dependent scattering at the zeroth site of the Wilson chain with on-site energies $\epsilon_{0 \downarrow} = -\epsilon_{0 \uparrow} > 0$
as seen from Eqs. (\ref{eq:HEffective1}) and (\ref{eq:effPara2}), affects the values of $\widetilde{B}_{\text{ELC}}$ and $T_K^{\text{ELC}}$. To this end, we interpret $\kappa$ as a free parameter of the effective
model. For the isotropic Kondo model in zero magnetic field it is known that ordinary (spin-independent) potential scattering can be approximately mapped to a modified electron DOS at the Fermi level or,
equivalently, to an effective coupling parameter $J_{\text{eff}}$. \cite{Krishnamurthy1980b} This approximation predicts that an increase of the scattering parameter reduces $J_{\text{eff}}$ and thus also the
Kondo temperature.

\begin{figure}
\centering{\includegraphics[width=1.0 \linewidth]{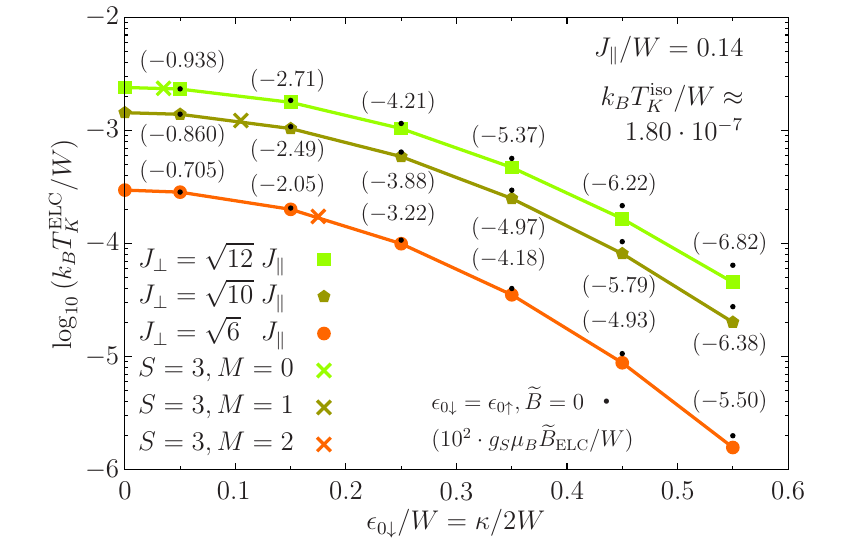}}
%
%
%
%
\caption{\label{fig:TKs_vs_ed}(Color online) Kondo temperature at the ELC field for the effective Hamiltonian with $g_e=0$ as function of the scattering parameter $\kappa$
(corresponding to a pair of on-site energies $\epsilon_{0 \downarrow} = -\epsilon_{0 \uparrow} > 0$ for the zeroth site of the Wilson chain) for fixed $J_{\parallel}$ and three values of the
coupling parameter $J_{\perp} > J_{\parallel}$. Numbers in parentheses denote the respective ELC field and lines are intended as a guide to the eye. Pessimistic error bars would be smaller
than the symbol size. Additional crosses indicate the Kondo temperature for parameters which, according to Eqs. (\ref{eq:effPara2}) and (\ref{eq:effPara3}), correspond to the given values of the
quantum numbers $S$ and $M$. Small dots mark $T_K$ for the case of potential scattering, i.e., for spin-independent on-site energies \emph{at zero magnetic field}.}
\end{figure}

In the example shown in Fig. \ref{fig:TKs_vs_ed}, spin quantum number $S=3$ (as in Fig. \ref{fig:FIKE_vs_B}) is considered and those coupling parameters $J_{\perp}$ are chosen which, as per Eq.
(\ref{eq:effPara3}), are assigned to the three magnetic quantum numbers allowed for this value of $S$ (i.e., $M=0,1,2$). Without scattering term, the Kondo temperature decreases when $J_{\perp}$
is reduced. \cite{Romeike2006a,Zitko2008b} Additional spin-dependent scattering further lowers $T_K^{\text{ELC}}$ just as standard potential scattering with $\epsilon_{0 \downarrow} = \epsilon_{0 \uparrow}$
at zero magnetic field does, but in comparison leads to smaller values of the Kondo temperature. In accordance with the expression for $J_{\text{eff}}$ from Ref. \onlinecite{Krishnamurthy1980b}, the
sign of the spin-\emph{independent} on-site energies does not affect $T_K$. Fig. \ref{fig:TKs_vs_ed} reveals that the decrease of $T_K^{\text{ELC}}$ accelerates with growing scattering strength.
Furthermore, we observe that the spin-dependent scattering has a larger influence on the Kondo temperature at the ELC field when the coupling parameter $J_{\perp}$ is smaller.

Let us now turn to the effect of $\kappa$ on the position of the ELC field. An additional spin-dependent scattering term breaks the spinflip-invariance and is therefore the very reason for a non-zero value of
$\widetilde{B}_{\text{ELC}}$. It thus seems plausible that a larger value of $\kappa$ also leads to a larger absolute value of the ELC field (cf. the numbers in parentheses in Fig. \ref{fig:TKs_vs_ed}). This
increase of $|\widetilde{B}_{\text{ELC}}|$ decelerates with growing scattering strength. A closer look at the data reveals that $\kappa$ again has a stronger effect when the coupling parameter $J_{\perp}$
is smaller.

Additional crosses in Fig. \ref{fig:TKs_vs_ed} mark the Kondo temperature for those values of the scattering parameter $\kappa$ that follow from Eq. (\ref{eq:effPara2}) for the three considered $M$
quantum numbers. We observe that the effective model predicts a decrease of $T_K^{\text{ELC}}$ with growing $M$, i.e., with every further ELC. As the example demonstrates, this decline of the
Kondo temperature is due to three cooperating effects: 1) According to Eq. (\ref{eq:effPara3}), $J_{\perp}$ becomes smaller when $M$ is increased. 2) Simultaneously, $\kappa$ becomes larger.
3) Because of the decreasing value of $J_{\perp}$, the scattering parameter additionally gains in importance. For the ELC fields $g_S \mu_B \widetilde{B}_{\text{ELC}}/W$ that belong to the three
special values of $T_K^{\text{ELC}}$, we obtain the following results (the error estimates indicate the variance with respect to $z$): $-6.59_{-0.25}^{+0.19} \cdot 10^{-3}$ $(S=3, M=0)$,
$-1.78_{-0.07}^{+0.06} \cdot 10^{-2}$ $(M=1)$, and $-2.36_{-0.09}^{+0.06} \cdot 10^{-2}$ $(M=2)$.


Finally, we investigate how the ELC field and the Kondo temperature at the ELC field depend on the parameters of the full Hamiltonian, i.e., on $J$, $S$, and $M$, with the parameters of
the effective model given by Eqs. (\ref{eq:effPara2}) to (\ref{eq:effPara4}). First of all, we note that all obtained (relative) ELC fields are negative. This supports the conclusion
that, in the impurity magnetization curves for equal g-factors and large hard axis anisotropy presented in Fig. \ref{fig:M_vs_BD-hardAxis}, the free LCs are only exceeded because of the
electrons' non-zero magnetic coupling. For $S=1$, $3/2$, $2$ and $J/W=0.14$ (as in Fig. \ref{fig:M_vs_BD-hardAxis}), the following values for $g_S \mu_B \widetilde{B}_{\text{ELC}}/W$
are obtained: $-3.90_{-0.14}^{+0.10} \cdot 10^{-3}$ ($S=1, M=0$), $-8.27_{-0.30}^{+0.22} \cdot 10^{-3}$ ($S=3/2, M=1/2$), $-4.95_{-0.18}^{+0.14} \cdot 10^{-3}$ ($S=2, M=0$),
$-1.31_{-0.05}^{+0.04} \cdot 10^{-2}$ ($S=2, M=1$).

\begin{figure}
\centering{\includegraphics[width=1.0 \linewidth]{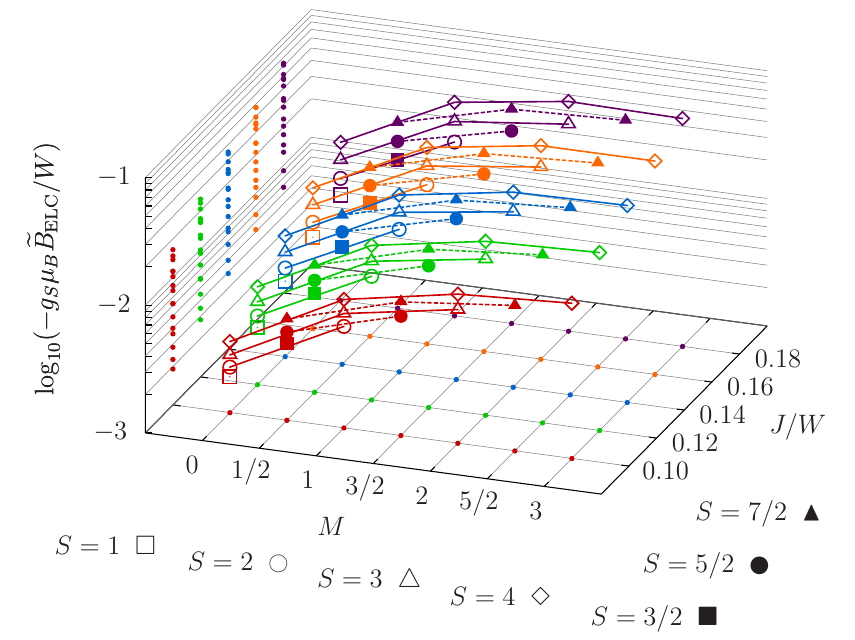}}
%
%
%
%
\caption{\label{fig:ELCs_BigBang}(Color online) Negative value of the ELC field for a set of parameter combinations of the effective Hamiltonian with $g_e=0$ which, according to Eqs.
(\ref{eq:effPara2}) to (\ref{eq:effPara4}), correspond to certain spin quantum numbers $S$, magnetic quantum numbers $0 \leq M < S$, and coupling parameters $J$. Projections of the
data points onto two planes are shown as small dots and lines connect points with the same value of $S$ and $J$. Pessimistic error bars would be smaller than the symbol size.}
\end{figure}

In contrast to the prediction of the effective model with $g_e>0$ from Eq. (\ref{eq:HEffective2}), the ELC fields depend on the quantum numbers $S$ and $M$ for vanishing electron
g-factor (see Fig. \ref{fig:ELCs_BigBang}). With increasing value of $S$, i.e., with increasing coupling parameter $J_{\perp}$, the absolute value of the ELC field grows as already seen
in Fig. \ref{fig:TKs_vs_ed}. A larger coupling $J$ increases all parameters of the effective model ($J_{\parallel}, J_{\perp}$, and $\kappa$) and, as demonstrated by Fig. \ref{fig:ELCs_BigBang},
thereby leads to a larger value of $|\widetilde{B}_{\text{ELC}}|$. It is furthermore evident that, with growing $J$, the quantum number $S$ gains in importance: Fig. \ref{fig:ELCs_BigBang}
shows that the ``curves'' for fixed coupling strength ``fan out'' for larger value of $J$. The dependence of the ELC field on the magnetic quantum number is the result of two counteracting
effects: 1) A larger value of $M$ leads to smaller coupling strength $J_{\perp}$ which, on its own, would lower $|\widetilde{B}_{\text{ELC}}|$. 2) On the other hand, the scattering
term becomes stronger with increasing $M$ and would, on its own, enlarge $|\widetilde{B}_{\text{ELC}}|$. For the parameters considered in Fig. \ref{fig:ELCs_BigBang}, there is a growth of
$|\widetilde{B}_{\text{ELC}}|$ with $M$ for $S \leq 7/2$ that decelerates with increasing $M$. In the case of $S=4$ and both $J/W=0.16$ and $J/W=0.18$, we observe a decrease of the
absolute value of the ELC field in the last step.


\begin{figure}
\centering{\includegraphics[width=1.0 \linewidth]{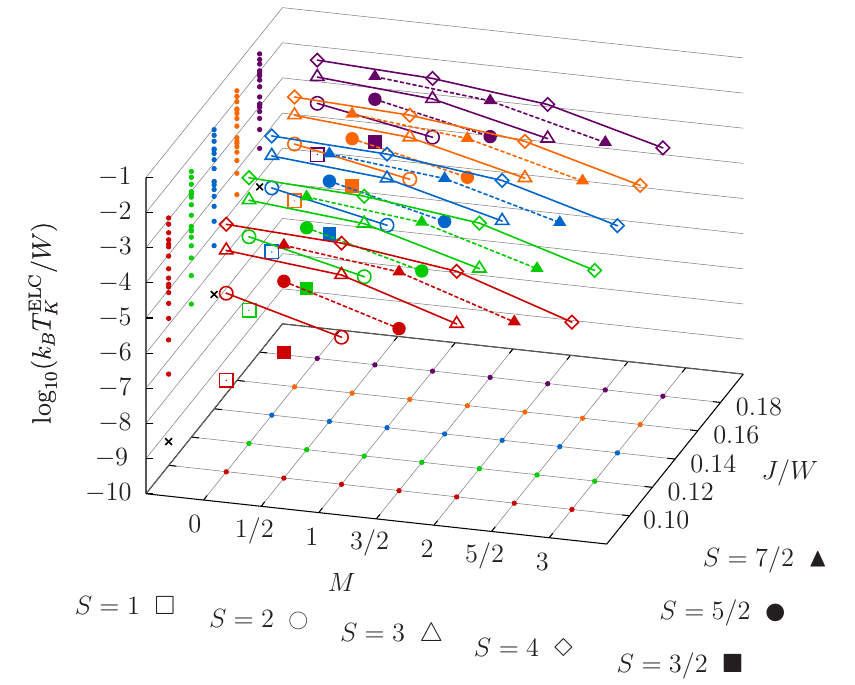}}
%
%
%
%
\caption{\label{fig:TKs_BigBang}(Color online) Kondo temperature at the respective ELC field for the same set of parameter combinations of the effective Hamiltonian with $g_e=0$ as in
Fig. \ref{fig:ELCs_BigBang}. Again, pessimistic error bars would be smaller than the symbol size. Three small crosses in the plane spanned by $k_BT_K^{\text{ELC}}/W$ and $J/W$ indicate
the value of the Kondo temperature for the isotropic case without scattering term (as inferred from Table \ref{tab:T_H-isotropic}).}
\end{figure}

To conclude this section, the Kondo temperatures belonging to the ELC fields shown in Fig. \ref{fig:ELCs_BigBang} are presented in Fig. \ref{fig:TKs_BigBang}. As the main result we find
that, according to the above explanation, the value of $T_K^{\text{ELC}}$ increasingly drops with growing $M$. On the other hand, a larger value of $S$ increases the coupling strength
$J_{\perp}$ and thus also $T_K^{\text{ELC}}$. For the parameters considered in Fig. \ref{fig:TKs_BigBang}, the ``last'' value of the Kondo temperature (i.e., for $M=S-1$ at fixed $J$) also
grows with $S$. It turns out that the influence of $S$ on $T_K^{\text{ELC}}$ is reduced for larger coupling $J$. As a consequence, the ``curves'' for fixed $J$ are ``focussed'' in the direction
of increasing coupling strength. We find that the dependence of the Kondo temperature on $J$ is the result of two counteracting effects: 1) Both coupling parameters $J_{\parallel}$ and
$J_{\perp}$ grow with $J$ and would, on their own, lead to a larger value of $T_K^{\text{ELC}}$. 2) However, the scattering parameter $\kappa$ is also increased and would, on its own,
lower the Kondo temperature. For all parameter combinations presented in Fig. \ref{fig:TKs_BigBang}, $T_K^{\text{ELC}}$ is a monotonously increasing function of the coupling strength
$J$. As a final observation, the relative decrease of the Kondo temperature between two consecutive values of $M$ (for fixed $J$) becomes smaller for larger quantum number $S$.

\subsubsection{\label{subsubsec:KondoEffectsComparison}Comparison of anisotropy- and field-induced pseudo-spin-1/2 Kondo effect for half-integer impurity spin}

In the case of half-integer impurity spin $S \geq 3/2$, two different types of pseudo-spin-1/2 Kondo effects occur for large hard axis anisotropy $D$. At zero magnetic field, the anisotropy
splits up the impurity multiplet and a doublet with magnetic quantum numbers $M = \pm 1/2$ remains as the lowest-lying impurity level (cf. Fig. \ref{fig:M_vs_B-hardAxis-freeSpin} b).
If the energy gap to the impurity states with $M = \pm 3/2$, which is equal to $2D$, is sufficiently large, this doublet undergoes spin-1/2 Kondo screening (let us call this Kondo effect
``anisotropy-induced''). \cite{Zitko2008b} At non-zero magnetic field, on the other hand, ELCs with associated ``field-induced'' Kondo effects occur, as discussed in the previous section.
Both types of Kondo effects show up in the corresponding impurity magnetization curves. In Fig. \ref{fig:M_vs_B-hardAxis-S3-2} a, for example, we observe the magnetic response of the
effective impurity doublet at low magnetic fields $g_S \mu_B B/D \ll 1$, which is then followed by a step at $g_S \mu_B B/D \approx 2$ due to the ELC.

One might wonder how the Kondo temperatures of the anisotropy- and field-induced Kondo effects compare. It has been shown in Ref. \onlinecite{Zitko2008b} that the anisotropy-induced Kondo
screening in the limit $D/W \rightarrow \infty$ is explained by Hamiltonian (\ref{eq:HEffective3}) with $\widetilde{B}=0$ and parameters $J_{\parallel} = J$, $J_{\perp}^D = (S+1/2)J$, and
$\kappa=0$. On the other hand, the pseudo-spin-1/2 Kondo effect at the first ELC (i.e., $M=1/2$) is described by the same Hamiltonian with $\widetilde{B} = \widetilde{B}_{\text{ELC}}$
and parameters $J_{\parallel} = J$, $J_{\perp}^{\text{ELC}} = \sqrt{(S-1/2)(S+3/2)} \, J$, and $\kappa = J$. In particular, we thus have $J_{\perp}^{\text{ELC}} < J_{\perp}^D$ for $S \geq 3/2$.
Since Fig. \ref{fig:TKs_vs_ed} shows that additional scattering $\kappa$ reduces the Kondo temperature, we can conclude that $T_K^D$ for the anisotropy-induced Kondo effect is always
larger than $T_K^{\text{ELC}}$ at the first ELC and, according to Fig. \ref{fig:TKs_BigBang}, at all following ELCs, too.

\section{\label{sec:summary}Summary}

In this article, we have reported on Numerical Renormalization Group (NRG) calculations for a Kondo model with additional uniaxial anisotropy $D$. Results have been presented for non-zero
magnetic field $B$ and different ratios $g_e/g_S$ of electron and impurity g-factor.

For a bulk field (i.e., for equal g-factors), a comparison of low-temperature NRG results for the impurity magnetization $\mathcal{M}(B)$ and the impurity contribution to the magnetization
$M_{\text{imp}}(B)$ reveals that $\mathcal{M} = \alpha M_{\text{imp}}$. The proportionality factor $\alpha > 1$ decreases for smaller couplings $\rho J_{\parallel}$ and $\rho J_{\perp}$.
Compared to the case of a local field (i.e., $g_e=0$), a non-zero electron g-factor effectively rescales the magnetic field argument of the impurity magnetization $\mathcal{M}(B)$ at low temperature.
Calculations for isotropic exchange interaction (i.e., $J_{\parallel} = J_{\perp} = J$) suggest that both effects must have a weak dependence on $D$ as long as the uniaxial anisotropy is small
compared to the bandwidth. In addition, the corresponding values of the rescaling factor $\eta(g_e/g_S=1)$ and the proportionality factor $\alpha$ apparently coincide.

For an isotropic impurity ($D=0$), we find that $\mathcal{M}(T \approx 0,B)$, unlike $M_{\text{imp}}$, does not display universal behavior in the usual sense as already noticed by Lowenstein.
\cite{Lowenstein1984} 

With additional easy axis anisotropy ($D<0$), the zero-temperature curve $M_{\text{imp}}(B)$ approaches a $D$-dependent value for small, but non-zero magnetic field, which corresponds to
the effective moment of the respective ``fractional spin'' \cite{Zitko2008b} as given by $T \chi_{\text{imp}}(B=0)$. The magnetic response at small fields and low temperature (relative to $|D|$) is
thus reminiscent of an ordinary magnetic doublet. Appropriately, the impurity magnetization $\mathcal{M}$ is well described by a rescaled and shifted Brillouin function in this regime.

In the case of hard axis anisotropy ($D>0$), a non-zero magnetic field can lead to ``effective level crossings'' (ELCs), at which pseudo-spin-1/2 Kondo screening occurs. For $g_e=0$ and
$D/W \rightarrow \infty$, these field-induced Kondo effects are described by an exchange-anisotropic spin-1/2 Kondo model with additional spin-dependent scattering at the zeroth site of
the Wilson chain. At the respective ELC field, this scattering leads to a reduction of the Kondo temperature $T_K^{\text{ELC}}$ in a similar way as ordinary potential scattering does at zero
magnetic field. In particular, the effective model predicts that $T_K^{\text{ELC}}$ decreases with every further ELC. This agrees with the observation that the steps in the magnetization
curves for large $D$, which are due to the field-induced Kondo effects, become steeper in the direction of increasing magnetic field.


\begin{acknowledgments}
We would like to thank T. Pruschke, S. Schmitt, and, in particular, T. Costi for valuable and helpful discussions.

Part of the calculations presented in this article have been done at the Leibniz-Rechenzentrum (LRZ) in Garching near Munich.

Financial support by the Deutsche Forschungsgemeinschaft (DFG) through research group FOR 945 is gratefully acknowledged.
\end{acknowledgments}


\appendix

\section{\label{app:NRGWithElectronZeeman}Numerical Renormalization Group calculations with conduction electron Zeeman term}

In this appendix, we briefly describe the changes to the standard NRG procedure \cite{Bulla2008} which are necessary in order to carry out calculations with an additional Zeeman term for the
conduction electrons.

\subsection{\label{subapp:discretization}Logarithmic discretization}

The starting point is the continuous energy representation of the Hamiltonian from Eq. (\ref{eq:energyRep}) with a restriction to the physically reasonable case $h < W$. In the following it is assumed
that the magnetic field, which appears in $h=g_e \mu_B B$, is non-zero and fixed and, to simplify the notation, that the chemical potential is zero. We introduce abbreviations for the absolute value
of the integration boundaries in Eq. (\ref{eq:energyRep}),

\begin{equation}
	\mathcal{B}^{\pm}_{\mu} = |\pm W + \mu h| \; ,
\end{equation}

\noindent rescale the integration variable $\varepsilon$, and change to rescaled electron operators (cf. Ref. \onlinecite{Krishnamurthy1980a}):

\begin{eqnarray}
	\xi_{\mu}^{+} & = & \frac{\varepsilon}{\mathcal{B}^{+}_{\mu}} \quad \text{for} \quad \varepsilon > 0 \; , \\
	\xi_{\mu}^{-} & = & \frac{\varepsilon}{\mathcal{B}^{-}_{\mu}} \quad \text{for} \quad \varepsilon < 0 \; , \\
	\op{a}_{\xi \mu}^{+} & = & \sqrt{\mathcal{B}^{+}_{\mu}} \, \op{a}_{\varepsilon \mu} \quad \text{for} \quad \varepsilon > 0 \; , \\
	\op{a}_{\xi \mu}^{-} & = & \sqrt{\mathcal{B}^{-}_{\mu}} \, \op{a}_{\varepsilon \mu} \quad \text{for} \quad \varepsilon < 0 \; .
\end{eqnarray}

\noindent Using $\int_{-W}^{W}{\mathrm{d} \varepsilon \, \rho(\varepsilon)} = 1$ and defining the normalized zeroth state of the Wilson chain as

\begin{eqnarray}
	\op{f}_{0 \mu} & = & \int_{0}^{1}{\mathrm{d} \xi_{\mu}^{+} \, \sqrt{\rho\left( \xi_{\mu}^{+} \mathcal{B}^{+}_{\mu} - \mu h \right) \mathcal{B}^{+}_{\mu}} \, \op{a}_{\xi \mu}^{+}} \\
	& + & \int_{-1}^{0}{\mathrm{d} \xi_{\mu}^{-} \, \sqrt{\rho \left(\xi_{\mu}^{-} \mathcal{B}^{-}_{\mu} - \mu h \right) \mathcal{B}^{-}_{\mu}} \, \op{a}_{\xi \mu}^{-}} \quad , \nonumber
\end{eqnarray}

\noindent we then obtain an equivalent expression for the electronic and interaction term in Eq. (\ref{eq:energyRep}):

\begin{eqnarray}
	\label{eq:rescaledH}
	\op{H}_{\text{cb+int}} & = & W \sum_{\mu}{\left( \frac{\mathcal{B}^{+}_{\mu}}{W} \int_{0}^{1}{\mathrm{d} \xi_{\mu}^{+} \, \xi_{\mu}^{+} \, \op{a}_{\xi \mu}^{+ \dagger} \op{a}_{\xi \mu}^{+} } \right.} \\
	& + & \left. \frac{\mathcal{B}^{-}_{\mu}}{W} \int_{-1}^{0}{\mathrm{d} \xi_{\mu}^{-} \, \xi_{\mu}^{-} \, \op{a}_{\xi \mu}^{- \dagger} \op{a}_{\xi \mu}^{-} } \right) \nonumber \\
	& + & J \, \opVec{S} \cdot \sum_{\mu, \nu}{\opDag{f}_{0 \mu} \frac{\bm{\sigma}_{\mu \nu}}{2} \op{f}_{0 \nu}} \quad . \nonumber
\end{eqnarray}

Next, the logarithmic discretization of the conduction band is carried out according to one of the available discretization schemes \cite{Yoshida1990, Oliveira1994, Costa1997, Campo2005,
Zitko2009a, Zitko2009b} by dividing the integration range $[-1,1]$ into standard intervals $I^{\pm}_{m}$ and using the following weight function on the $m^{\text{th}}$ positive and negative
interval, respectively:

\begin{equation}
	\varphi^{\pm}_{m \mu}(\xi_{\mu}^{\pm}) = \sqrt{\frac{\rho \left(\xi_{\mu}^{\pm} \mathcal{B}^{\pm}_{\mu} - \mu h \right)}{\int_{I^{\pm}_m}{\mathrm{d} \xi_{\mu}^{' \pm} \, \rho \left( \xi_{\mu}^{' \pm} \mathcal{B}^{\pm}_{\mu} - \mu h \right)}}} \quad .
	\label{eq:weightFunction}
\end{equation}

\noindent With $s = \pm$,

\begin{equation}
	\gamma^{s}_{m \mu} = \sqrt{\frac{\mathcal{B}^{s}_{\mu}}{W} \int_{I^{s}_m}{\mathrm{d} \xi^{s}_{\mu} \, \rho \left( \xi^{s}_{\mu} \mathcal{B}^{s}_{\mu} - \mu h \right) W}} \quad ,
\end{equation}

\noindent and new operators $\op{a}^{s}_{m \mu}$ corresponding to the weight functions $\varphi^s_{m \mu}(\xi_{\mu}^s)$ on the intervals $I_m^s$, we have the following exact expansion
for the zeroth state of the Wilson chain:

\begin{equation}
	\op{f}_{0 \mu} = \sum_{s, m}{\gamma^{s}_{m \mu} \, \op{a}^{s}_{m \mu}} \; \; .
\end{equation}

\noindent In addition, a dimensionless ``energy'' $\mathcal{E}^{s}_{m \mu}$ has to be assigned to each interval $I^{s}_{m}$ for each spin projection $\mu$. This is done according to the chosen
discretization scheme by using the weight function (\ref{eq:weightFunction}) with the shifted DOS, leading to a discrete approximation to Hamiltonian (\ref{eq:rescaledH}):

\begin{eqnarray}
	\label{eq:HDiscrete}
	\op{H}_{\text{cb+int}} & \rightarrow & W \sum_{s, m, \mu}{\frac{\mathcal{B}^{s}_{\mu}}{W} \, \mathcal{E}^{s}_{m \mu} \, \op{a}^{s \dagger}_{m \mu} \, \op{a}^{s}_{m \mu} } \\
	& + & J \, \opVec{S} \cdot \sum_{\mu, \nu}{\opDag{f}_{0 \mu} \frac{\bm{\sigma}_{\mu \nu}}{2} \op{f}_{0 \nu}} \quad . \nonumber
\end{eqnarray}

At this point, the substitution (\ref{eq:HDiscrete}) is still valid for arbitrary $\rho(\varepsilon)$. The above expressions simplify in the case of a constant density of states, $\rho(\varepsilon) = 1/2W$, as a
shifted constant DOS is, of course, still a constant DOS:

\begin{eqnarray}
	\op{H}_{\text{cb+int}} & \rightarrow & W \sum_{s, m, \mu}{\frac{\mathcal{B}^{s}_{\mu}}{W} \underbrace{\mathcal{E}^{s}_{m \mu}(h=0)}_{= \; \mathcal{E}^{s}_{m}} \, \op{a}^{s \dagger}_{m \mu} \, \op{a}^{s}_{m \mu} } \nonumber \\
	& + & J \, \opVec{S} \cdot \sum_{\mu, \nu}{\opDag{f}_{0 \mu} \frac{\bm{\sigma}_{\mu \nu}}{2} \op{f}_{0 \nu}} \; \; , \label{eq:HDiscreteSimple} \\
	\op{f}_{0 \mu} & = & \sum_{s, m}{\sqrt{\frac{\mathcal{B}^{s}_{\mu}}{W}} \underbrace{\gamma^{s}_{m \mu}(h=0)}_{= \; \gamma_{m}} \, \op{a}^{s}_{m \mu}} \; \; . \label{eq:WilsonStateSimple}
\end{eqnarray}

\noindent Here, $\mathcal{E}^{s}_{m}$  and $\gamma_{m}$ are the ``energies'' and expansion coefficients, respectively, for the system with a \emph{local} magnetic field (i.e., with $g_e = 0$).

\subsection{\label{subapp:tridiagonalization}Tridiagonalization}

Since the rescaling factors

\begin{equation}
	\frac{\mathcal{B}^{s}_{\mu}}{W} = \left| s + \mu \frac{g_e}{g_S} \frac{g_S \mu_B B}{W} \right|
	\label{eq:rescalingFactor}
\end{equation}

\noindent depend on spin projection $\mu$ and magnetic field $B$, the tridiagonalization of Hamiltonian (\ref{eq:HDiscreteSimple}), which leads to the Wilson chain with hopping parameters
$t_{i \mu}(B)$ and on-site energies $\epsilon_{i \mu}(B)$, has to be done separately for spin-up and spin-down and for each value of $B$. In case of a constant DOS, Eqs.
(\ref{eq:HDiscreteSimple}) and (\ref{eq:WilsonStateSimple}) show that the only necessary modification of an existing code solving the recursion relations given in Ref. \onlinecite{Bulla2008}
is to multiply all ``energies'' $\mathcal{E}^{s}_{m}$ and coefficients $\gamma_{m}^2$ with the appropriate factor (\ref{eq:rescalingFactor}).

For a particle-hole symmetric DOS we have $\mathcal{E}^{s}_{m \mu} = -\mathcal{E}^{-s}_{m -\mu}$ and $\gamma^{s}_{m \mu} = \gamma^{-s}_{m -\mu}$. Using the Ansatz
$u_{nm \mu} = (-1)^n v_{nm -\mu}$ and $v_{nm \mu} = (-1)^n u_{nm -\mu}$ for the coefficients of the orthogonal transformation (following the notation of Ref. \onlinecite{Bulla2008}),
it can then be shown that $t_{i \uparrow}(B) = t_{i \downarrow}(B)$ and $\epsilon_{i \uparrow}(B) = -\epsilon_{i \downarrow}(B)$ for all sites $i$ of the Wilson chain.



%


\end{document}